\documentclass{aa}
\usepackage{amsmath, amssymb, graphics}
\usepackage{hyperref}
\usepackage{epstopdf}
\usepackage{natbib}
\begin{document}

\title{Orbital Stability of Earth Trojans}

\author{Lei Zhou\inst{1} \and Yang-Bo Xu\inst{1} \and Li-Yong Zhou\inst{1,2} \and  Rudolf Dvorak\inst{3} \and Jian Li\inst{1,2}}
\authorrunning{Zhou et al}
\offprints{L.-Y. Zhou, \email zhouly@nju.edu.cn}

\institute{School of Astronomy and Space Science, Nanjing
	University, 163 Xianlin Avenue, Nanjing 210046, China \and Key Laboratory of Modern
	Astronomy and Astrophysics in Ministry of Education, Nanjing
	University, Nanjing 210046, China \and Universit\"{a}tssternwarte Wien, T\"{u}rkenschanzstr. 17, 1180 Wien, Austria}

\date{}

\abstract{The only discovery of Earth Trojan 2010 TK$_7$ and the subsequent launch of OSIRIS-REx motive us to investigate the stability around the triangular Lagrange points $L_4$ and $L_5$ of the Earth. In this paper we present detailed dynamical maps on the $(a_0,i_0)$ plane with the spectral number (SN) indicating the stability. Two main stability regions, separated by a chaotic region arising from the $\nu_3$ and $\nu_4$ secular resonances, are found at low ($i_0\leq 15^\circ$) and moderate ($24^\circ\leq {i_0}\leq 37^\circ$) inclinations respectively. 
The most stable orbits reside below $i_0=10^\circ$ and they can survive the age of the Solar System.
The nodal secular resonance $\nu_{13}$ could vary the inclinations from $0^\circ$ to $\sim 10^\circ$ according to their initial values while $\nu_{14}$ could pump up the inclinations to $\sim 20^\circ$ and upwards. The fine structures in the dynamical maps are related to higher-degree secular resonances, of which different types dominate different areas. The dynamical behaviour of the tadpole and horseshoe orbits, reflected in their secular precession, show great differences in the frequency space. The secular resonances involving the tadpole orbits are more sensitive to the frequency drift of the inner planets, thus the instabilities could sweep across the phase space, leading to the clearance of tadpole orbits. We are more likely to find terrestrial companions on horseshoe orbits.
The Yarkovsky effect could destabilize Earth Trojans in varying degrees. We numerically obtain the formula describing the stabilities affected by the Yarkovsky effect and find the asymmetry between the prograde and retrograde rotating Earth Trojans. The existence of small primordial Earth Trojans that avoid being detected but survive the Yarkovsky effect for 4.5\,Gyr is substantially ruled out.  }


\maketitle

\section{Introduction}\label{sec:intro}

{}

In the circular restricted three-body model for a massless asteroid moving in the gravitational field of the Sun and a planet, the equilateral triangular Lagrange equilibrium points, usually denoted by $L_4$ and $L_5$ for the leading and trailing one respectively, are dynamically stable for all planets in our Solar System \citep[see e.g.][]{1999ssd..book.....M} and small celestial objects may find stable residence around them. The first such asteroid librating around the $L_4$ point of Jupiter (588 Achilles) was discovered in 1906 by \citet{1907AN....174...47W}, nowadays several thousand objects are observed orbiting around either $L_4$ or $L_5$ point of the Earth, Mars, Jupiter, Uranus and Neptune (see lists at IAU: Minor Planet Center (MPC) \footnote{\url{https://www.minorplanetcenter.net/iau/lists/Trojans.html}}). These asteroids are called \textit{Trojans} after the mythological story of Trojan War. A Trojan asteroid shares the same orbit with its parent planet and in fact is locked in the 1:1 mean motion resonance (MMR) with the planet.

Many studies have been devoted to the Trojan dynamics \citep[e.g.][]{1990AJ....100..290M,1992AJ....104.1641M,2002Icar..160..271N,2004LPI....35.1107S,2009MNRAS.398.1217Z,2011MNRAS.410.1849Z,2009MNRAS.398.1715L,2011MNRAS.412..537L,2012MNRAS.426.3051C}. This topic is of special interest not only because a Trojan may exhibit complicated orbital behavior, but also because the existence and properties of Trojans are the touchstone by which the realness of the scenarios proposed for the early evolution of our Solar System \citep{2005Natur.435..462M,2009AJ....137.5003N} can be checked.

As for the Earth, observational searches for Trojans face unique difficulties due to the particular viewing geometry. The nearness of Trojans to the Earth leads to the wide area of sky to be searched. On the other hand, their locations close to the triangular Lagrange points place them in the daytime sky making the observations suffer higher airmass and the increased sky brightness of twilight \citep{2000Icar..145...33W}.
In 2010, a $300\;{\rm m}$ object ($2010~{\rm TK_7}$) was detected by the Wide-field Infrared Survey Explorer (WISE) \citep{2011ApJ...731...53M} and subsequently was confirmed to be the first Earth Trojan \citep{2011Natur.475..481C}. At the same time, another candidate $2010~{\rm SO_{16}}$ was identified as a horseshoe companion of the Earth \citep{2011MNRAS.414.2965C}.

The asteroid  $2010~{\rm TK_7}$ lies in an eccentric and inclined orbit ($e \sim0.19$, $i\sim 21^\circ$) around the $L_4$ point of the Earth. For its orbital elements, see e.g. \textit{Asteroids-Dynamic Site} (AstDyS-2)\footnote{\url{http://hamilton.dm.unipi.it/astdys/}}. \citet{2012A&A...541A.127D} have demonstrated that the asteroid $2010~{\rm TK_7}$ is situated out of the stability zone, leading to a total lifetime of being in the 1:1 MMR with the Earth less than 0.25 Myr. They also performed a numerical investigation of the phase space within a truncated planetary system from Venus to Saturn (Ve2Sa) and constructed a stability diagram for Earth Trojans adopting the maximum eccentricity as the stability indicator. They suggested that the apsidal secular resonances with Venus, the Earth, Mars and Jupiter can be responsible for the structure of the dynamical map. However, the perturbations from Uranus and Neptune could make some difference to the dynamical behavior of Earth Trojans and the integration time of 10 Myr may not be long enough for the maximum eccentricity to indicate the long term stability.

\citet{2000MNRAS.319...63T} carried out numerical surveys on the fictitious Earth Trojans with a complete model including all of the planets in our Solar System and found test particles can retain stable orbits at low ($i \la 16^\circ$) and moderate ($24^\circ \la {i} \la 34^\circ$) inclinations with their semi-major axis extending to $1\pm~0.012~{\rm AU}$. According to the investigations of \citet{2012A&A...541A.127D}, the corresponding stability windows for inclinations are $i \la 20^\circ$ and $28^\circ \la {i} \la 40^\circ$ respectively. Moreover, they found a small U-shaped stability region around $i=50^\circ$ (see Fig.~8 in \citealt{2012A&A...541A.127D}). \citet{2013CeMDA.117...91M} took the Yarkovsky force into account and purported that the most stable Earth Trojans can survive for a few Gyr, which means there could be primordial Trojans to be found in the tadpole regions of the Earth.

This paper proffers a more detailed dynamical map of Earth Trojans and explores the significant resonances that carve the stability diagram. In Section 2, we introduce the dynamical model and numerical algorithm. In addition, we illustrate the method of the spectral analysis and explain how we obtain the proper frequency of the fictitious Earth Trojans. In Section 3, we show the dynamical maps on the plane of $(a_0,i_0)$. With the help of them, we then determine the possible stability regions for Earth Trojans. In Section 4, we employ a frequency analysis method and derive the resonances that are responsible for the structures of the phase space. The influences of the Yarkovsky effect on Earth Trojans are discussed in Section 5. Finally, Section 6 presents the conclusions and discussions.

\section{Model and Method}\label{sec:mod}

\subsection{Dynamical model and initial conditions}\label{subsec:dymod}

To investigate the orbital stability and dynamical behaviour of Earth Trojans, we numerically simulated their orbital evolution and then assessed their orbital stability by a method of spectral analysis. The dynamical model adopted in our simulations, which is referred to as Ve2Ne hereafter, is consisted of the Sun, all of the planets in our Solar System except Mercury (from Venus to Neptune) and massless fictitious Earth Trojans (test particles). Mercury is excluded because it has negligible influence on the evolution of Earth Trojans in this research while the model including Mercury consumes much longer computation time \citep{2012A&A...541A.127D}. We adopt the Earth-Moon barycenter instead of the separate Earth and Moon as \citet{2012A&A...541A.127D} did. The initial orbits of the planets at epoch of JD 245\,7400.5 are derived from the JPL HORIZONS system\footnote{\url{ssd.jpl.nasa.gov/horizons.cgi}} \citep{1996DPS....28.2504G}.

We initialize the orbital elements of the fictitious Trojans in a similar way to \citet{2012A&A...541A.127D}. The test particles share the same eccentricity $e$, longitude of the ascending node $\Omega$ and mean anomaly $M$ with the Earth. The argument of perihelion $\omega_0$ is set as $\omega_0=\omega_3\pm60^\circ$, where the triangular Lagrange points lie\footnote{The subscript `2' to `8' denotes planet Venus to Neptune respectively throughout this paper.}. We sample the initial semi-major axes and inclinations uniformly on the $(a_0,i_0)$ plane in order to reveal their correlation with the stability. The semi-major axis ranges from 0.99 to 1.01 AU with an interval of $10^{-4}~{\rm AU}$ and
the inclinations are varied from $0^\circ$ to $60^\circ$ in a step of $1^\circ$.

The Yarkovsky effect acting on a rotating body is a radiation force caused by the anisotropic thermal re-emission. Small objects, especially meteoroids and small asteroids will undergo a semi-major axis drift under the perturbation of the Yarkovsky effect \citep{1951PRIA...54..165O}. We performed an extra set of simulations to explore how the Yarkovsky effect influences the dynamical behaviour of Earth Trojans. We adopt a complete linear model proposed by \citet{1999A&A...344..362V} to simulate the Yarkovsky effect in our calculations.

We implement a Lie-series integrator \citep{1984A&A...132..203H} to integrate the whole system. The \textit{hybrid symplectic integrator} in the \textsc{mercury6} software package  \citep{1999MNRAS.304..793C} are also implemented for simulations including the Yarkovsky effect.

\subsection{Spectral analysis}\label{subsec:spe}

A spectral analysis method is applied in our orbital integrations to remove the short-period terms and reduce the amount of the output data. Furthermore, the method provides us an accurate stability indicator to construct the dynamical map in Section~\ref{subsec:dynmap}.

We employ an on-line low-pass digital filter \citep{1993CeMDA..56..121M,1995A&A...303..945M} in smoothing the output of the preliminary integration, of which the interval is chosen to be 16 days. We then resample the filtered data with an interval of $\Delta=32,768~{\rm days}\ (\approx90~{\rm yr})$.
The whole system is integrated for $\sim1.2\times10^7~{\rm yr}$ so that we obtain $N=2^{17}\ (=131,072)$ lines of signal in time domain for further analysis.

A fast Fourier transform (FFT) is applied to the filtered data afterwards. The corresponding Nyquist frequency is $f_{\rm Nyq}=1/(2\Delta)=5.557\times10^{-3}~{\rm yr}^{-1}$, which is larger than all the fundamental secular frequencies (see Section~\ref{subsec:dyspe}) in our Solar System. The spectral resolution of the filtered data is $f_{\rm res}={1}/(N\Delta)=8.504\times10^{-8}~{\rm yr}^{-1}$.

The \textit{spectral number} (SN) is defined to be the number of the peaks over a specific threshold in a frequency spectrum \citep{2002Icar..158..343M,1995A&A...303..945M,2009MNRAS.398.1217Z,2011MNRAS.410.1849Z}. The spectral number could reflect the long term stability within a relatively short integration time. We make use of the spectral number as stability indicator in our dynamical maps as \citet{2009MNRAS.398.1217Z,2011MNRAS.410.1849Z} did for Neptune Trojans.

The critical angle (resonant angle) for a Trojan in the 1:1 MMR with the Earth is $\sigma=\lambda-\lambda_3$, where $\lambda=\omega+\Omega+M$ is the mean longitude. In this paper, we mainly use the spectral number of $\cos{\sigma}$ to construct the dynamical maps.

\subsection{Numerical analysis of proper frequencies}\label{subsec:naff}

\citet{1990Icar...88..266L} introduced a method based on the evolution of the proper frequencies with time to analyze the stability in a conservative dynamical system. This so-called ``Frequency Map Analysis (FMA)'' relies heavily on the accuracy of the determination of the proper frequency. A refined numerical algorithm that is several orders of magnitude more precise than the simple FFT was applied in FMA \citep{1990Icar...88..266L,1992PhyD...56..253L,1993PhyD...67..257L,1993CeMDA..56..191L}. For the high precision and feasibility, we implement the same algorithm to define the proper frequencies of Earth Trojans. Here we just outline the numerical algorithm.

Any given quasi-periodic function $f(t)$ in the complex domain can be expressed in the form

\begin{equation}
  f(t)=\sum_{k=1}^{\infty}a_k e^{\mathrm{i}\omega_k{t}}\,,
\end{equation}
where $a_k$ are the complex amplitudes (in descending order) of the corresponding periodic terms dominated by the frequencies $\omega_k$. This algorithm can numerically provide a precise recovery of $f(t)$ over a finite time span $[-T,~T]$:

\begin{equation}
  \widetilde{f}(t)=\sum_{k=1}^{N}\widetilde{a}_k e^{\mathrm{i}\widetilde{\omega}_k{t}}\,.
\end{equation}

The frequencies and amplitudes can be determined by an iterative scheme. At first, we conduct a modified FFT to $f(t)$:

\begin{equation}\label{equ:fft}
  \Psi(\omega)=\frac{1}{2T}\int_{-T}^{T}\!f(t)e^{-\mathrm{i}\omega{t}}\chi(t)\,dt\,,
\end{equation}
where $\chi(t)$ is a weight function that satisfies

\begin{equation}
  \frac{1}{2T}\int_{-T}^{T}\chi(t)\,dt=1\,.
\end{equation}
As always, we use the Hanning window $\chi(t)=1+\cos(\pi{t}/T)$ as the weight function for reducing the aliasing. $\widetilde{\omega}_1$ can be estimated by searching for the maximum term of the amplitude function $\Psi(\omega)$. Then we refine the estimation of $\widetilde{\omega}_1$ in its neighbourhood by maximizing Eq.~\eqref{equ:fft} and the amplitude $\widetilde{a}_1$ can be derived by orthogonal projection on $e^{\mathrm{i}\widetilde{\omega}_1{t}}$.

The above steps are repeated on the new function $f_1(t)=f(t)-\widetilde{a}_1 e^{\mathrm{i}\widetilde{\omega}_1{t}}$ for the following frequency. Last but not least, we have to orthogonalize the basis $e^{\mathrm{i}\widetilde{\omega}_k{t}}$ every time a new frequency is determined.

The iteration can be stopped either when the desired number of the frequencies is reached, or when the amplitude associated with the last frequency falls below a specific noise level.

\section{Dynamical map}\label{sec:res}

Once the numerical integration is completed, an FFT is conducted to the output. From the frequency spectrum of $\cos\sigma$, we obtain the SN of each orbit on the $(a_0,i_0)$ plane by counting the number of peaks over 1 per cent of the highest one. As mentioned before, the smaller the SN is, the more regular the orbit is.

\subsection{Dynamical map}\label{subsec:dynmap}
\begin{figure}[htbp]
	\centering
  \includegraphics[scale=0.29]{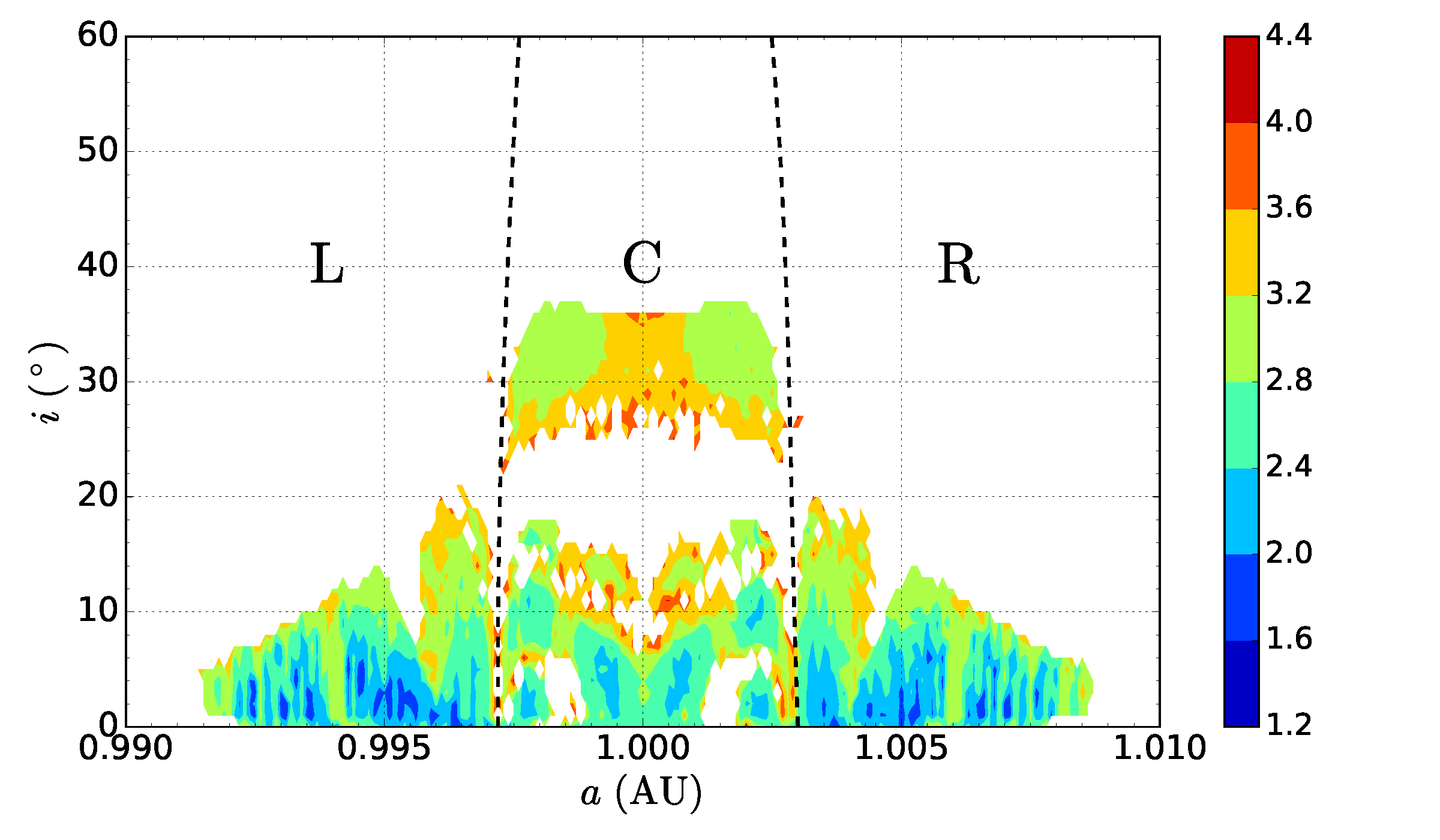}
  \caption{The dynamical map around the $L_4$ point on the $(a_0,i_0)$ plane. The colour indicates the SN of $\cos{\sigma}$, which is displayed on a base-10 logarithmic scale for more details. The orbits that escape from the Earth co-orbital region (see text) during the integration time (12\,Myr) are excluded. The dashed lines indicate the separatrices between the tadpole and horseshoe orbits (see text in Section~\ref{subsec:tadhorse}), which divide the dynamical map into three regimes. The tadpole orbits reside in the central region (denoted by ``C'') while the horseshoe orbits could be found in the left (``L'') and right (``R'') regions.}
  \label{fig:SNL4}
\end{figure}

We present the dynamical map around the $L_4$ point on the $(a_0,i_0)$ plane in Fig.~\ref{fig:SNL4}. The colour represents the base-10 logarithm of SN, which now serves as an indicator of the orbital stability. Orbits in blue are of the greatest stability while those in red are very close to the chaos. The orbits that dissatisfy $0.98\,{\rm AU}\leq{a}\leq1.02\,{\rm AU}$ at any time in the integration are regarded as escaping from the 1:1 MMR and they are excluded in the dynamical maps. The criterion is derived empirically by inspecting the orbital evolution of Earth Trojans.

\begin{figure}
	\centering
  \includegraphics[scale=0.29]{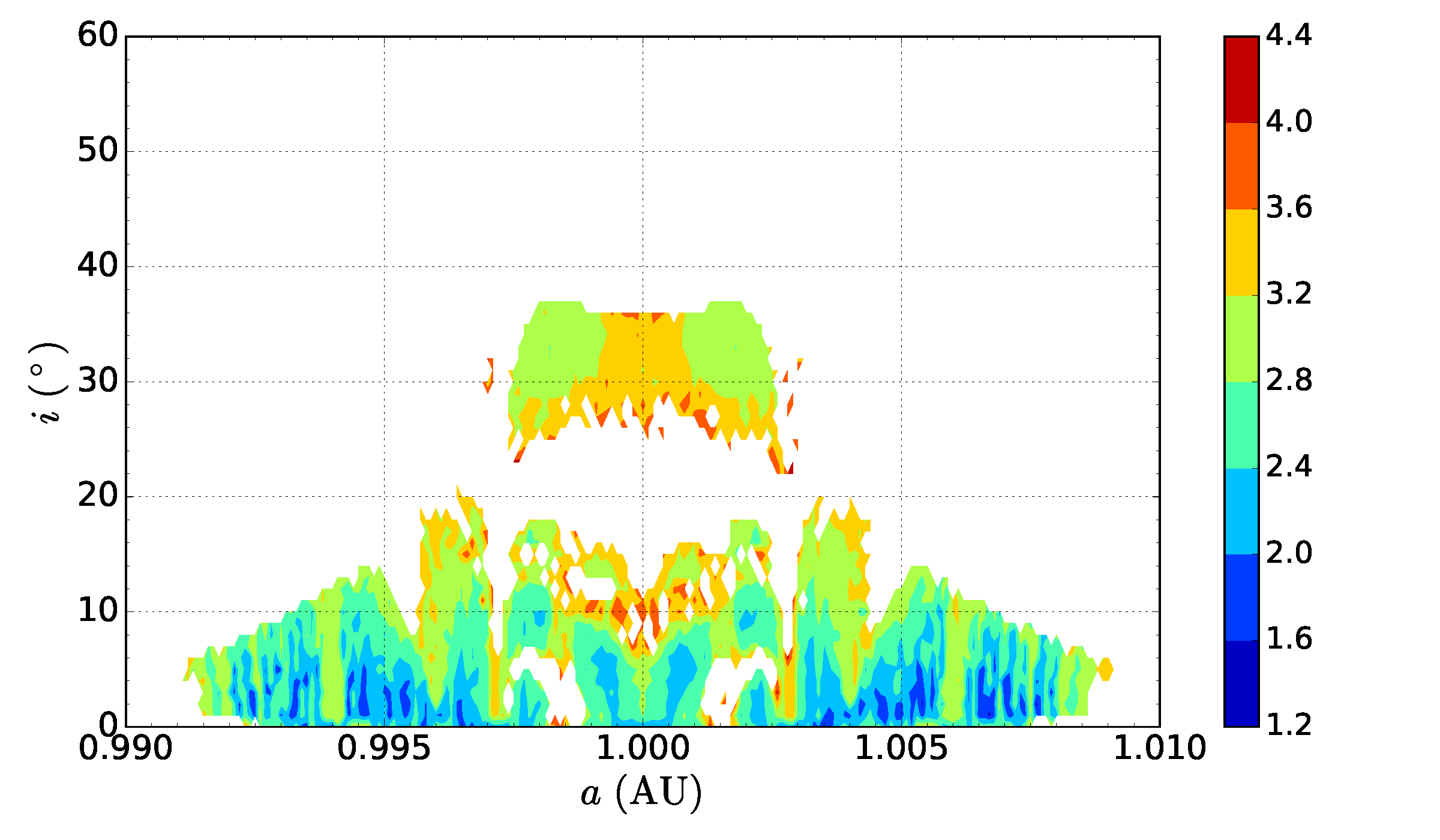}
  \caption{The same as Figure~\ref{fig:SNL4} but for $L_5$ point.}
  \label{fig:SNL5}
\end{figure}

As can be seen from Fig.~\ref{fig:SNL4}, the stability regions show an apparent symmetry about $a_0\approx 1$\,AU. For low inclinations, the stability window extends to $\pm0.0085$\,AU for semi-major axis centered at 1\,AU, which is a bit broader than that of the coplanar orbits. Then the region shrinks with the increasing inclinations until $i_0\approx{15^\circ}$, where an instability strip appears. For inclinations larger than $24^\circ$, a gamepad-shaped island spanning a range of 0.9975--1.0025 AU for semi-major axis is found to be able to hold stable orbits. In the region above $37^\circ$, no orbit can survive for 12\,Myr as a co-orbiting companion of the Earth. Another point to note is that the stability regions are being eroded by instabilities. Two rifts stand at $1\pm 0.0028$\,AU, and serve to divide the dynamical map into three regimes. In the central area, a pair of white carves indicating escaped orbits lie around $\pm 0.0015$\,AU about the centre while a V-shaped instability barrier is settled above them around $i_0\approx10^\circ$.
Besides, at the bottom of the stability island around $i_0=30^\circ$, some orbits are excited to give rise to the instability.

Left and right areas are symmetrical about $a_0\approx 1$\,AU and in each of them, there exists a triangular gap at the boundary of the stability region. The orbits surrounded by these instability strips could be endowed with SN over $10^4$ (in red) and they will escape from the Earth co-orbital region in the near future.

The most stable orbits of which the SN is smaller than 100 mainly reside in the region of $i_0\leq10^\circ$. A further simulation up to 4.5\,Gyr reveals that these orbits could survive the age of the Solar System. We may most possibly observe Earth Trojans in slightly inclined orbits because they occupy the largest stable area in the phase space according to Fig.~\ref{fig:SNL4}.

It is already known that phase spaces around the $L_4$ and $L_5$ points are dynamically identical to each other \citep[see e.g.][]{2009MNRAS.398.1217Z}. We calculated the dynamical map around the $L_5$ point and present it in Fig.~\ref{fig:SNL5}. It is almost the same as the one in Fig.~\ref{fig:SNL4}, and no remarkable dynamical asymmetry between the $L_4$ and $L_5$ points can be found. As a result, we could investigate only one of them then the same conclusion can be promoted to the other. In this paper we will focus on the $L_4$ point.

%Secular resonances such as $\nu_3$ and $\nu_5$ should play a vital role in carving the dynamical map (see Section~\ref{subsec:secres}). In addition, the overlap of high-degree secular resonances could bring in chaos.

\subsection{Region around $i=50^{\circ}$}\label{subsec:50d}

At high inclinations ($i\gtrsim40^\circ$), Earth Trojans will be trapped in Kozai mechanism \citep{1962AJ.....67..591K,1962P&SS....9..719L}. As a result of the increasing eccentricity, the Trojans will sustain close encounters with the planets, which give rise to the instability \citep{2004CeMDA..88..123B}. For Earth Trojans at moderate inclinations, the $\nu_3$ and $\nu_4$ secular resonances may affect their orbits and increase the eccentricity \citep{2002MNRAS.334..241B}.
Note that in accordance with practice, we denote the secular resonance as $\nu_i$ when $g=g_i$ and $\nu_{1i}$ when $s=s_i$, where $g$ ($g_i$) and $s$ ($s_i$) represent the precession rate of the perihelion and ascending node of Trojans (planet).

\begin{figure}[htbp]
		\centering
	\includegraphics[scale=0.26]{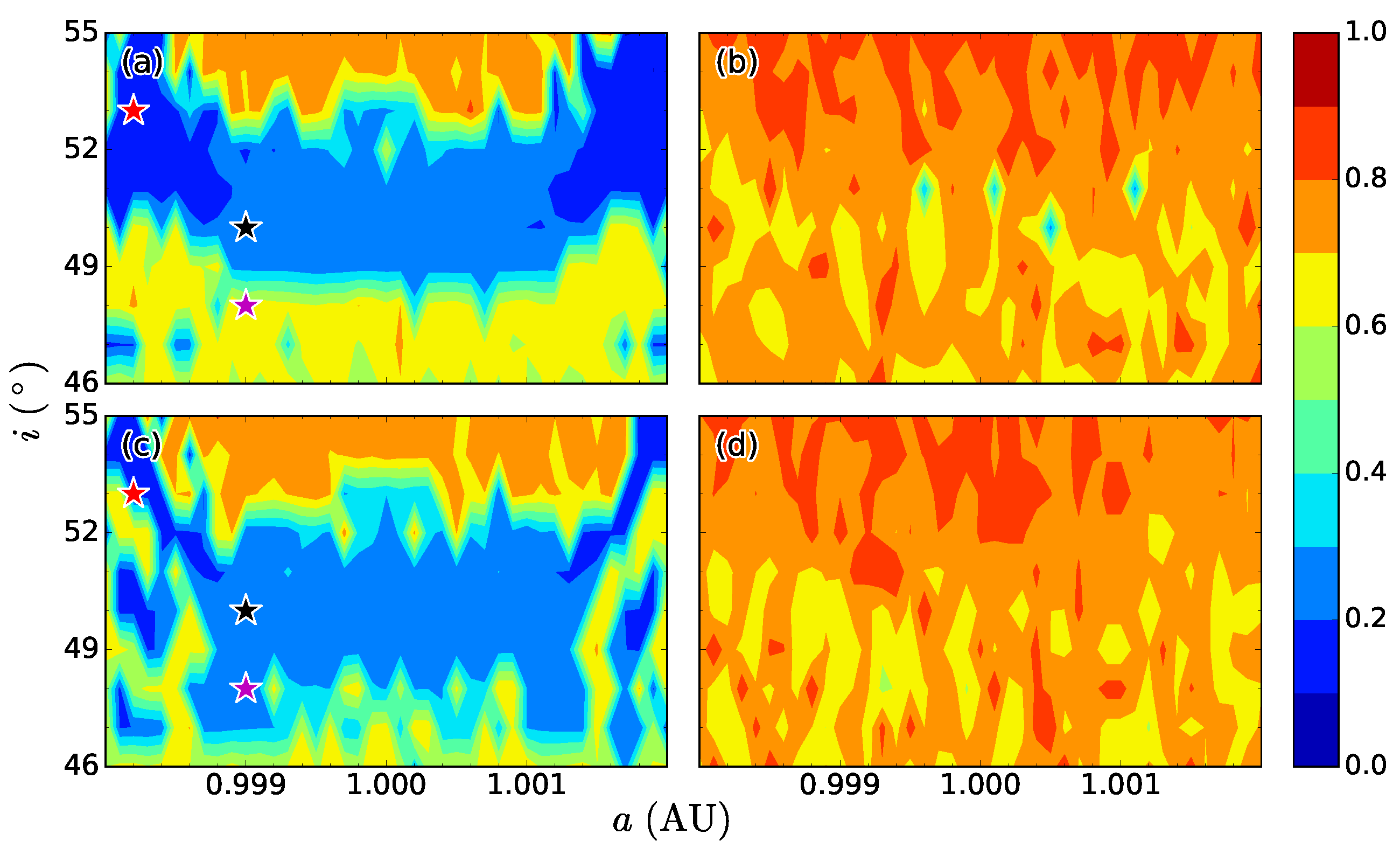}
	\caption{The dynamical maps for the possible stability window around $i=50^\circ$ \citep{2012A&A...541A.127D} on the $(a_0,i_0)$ plane. The colour indicates the maximum eccentricity during the integration time. The four panels stand for the results derived from different models and integration time, (a) Ve2Sa and integrated for 1\,Myr, (b) Ve2Sa for 12\,Myr, (c) Ve2Ne for 1\,Myr, and (d) Ve2Ne for 12\,Myr. The stars in panels (a) and (c) indicate the initial conditions for the orbits displayed in Fig.~\ref{fig:50Dorb}. Magenta, black and red respectively stand for $(a_0,i_0)=(0.9990,48^\circ)$, $(0.9990,50^\circ)$ and $(0.9982,53^\circ)$.}
	\label{fig:50D}
\end{figure}

\citet{2012A&A...541A.127D} adopted the maximum eccentricity as the stability indicator and constructed a similar stability diagram. In that investigation, a small stability region appears around $i=50^\circ$, which is unexpected according to our simulations. Considering the shorter integration time as well as the absence of Uranus and Neptune in the model Ve2Sa, we have to check the results simulated with different models and for different integration time to verify the existence of the aforementioned stability region. We run 4 sets of simulations with different combinations of the model (Ve2Sa and Ve2Ne) and integration time (1\,Myr and 12\,Myr). The results are summarized in Fig.~\ref{fig:50D}.

The left two panels in Fig.~\ref{fig:50D} confirm that, for integration time of 1\,Myr, the stability window around $i=50^\circ$ does exist for both models. However, this stability structure will shrink as the integration time goes on and then completely disappear within 12\,Myr (see panels (b) and (d) in Fig.~\ref{fig:50D}).

The orbits in the blue regions, whose edges look quiet fragmentized, are supposed to possess a maximum eccentricity smaller than 0.3 within 1\,Myr. Among these orbits, the most stable ones with $e_{\max}\le 0.2$ are at both wings of the blue areas. As we can see from panels (a) and (c), it seems that the inclusion of Uranus and Neptune will not prevent the formation of the stability window, but shape its structure. For the model containing Uranus and Neptune, the stability region extends to lower inclinations while its both sides are reduced.

\begin{figure*}[htbp]
	\centering
	\includegraphics[width=0.32\textwidth]{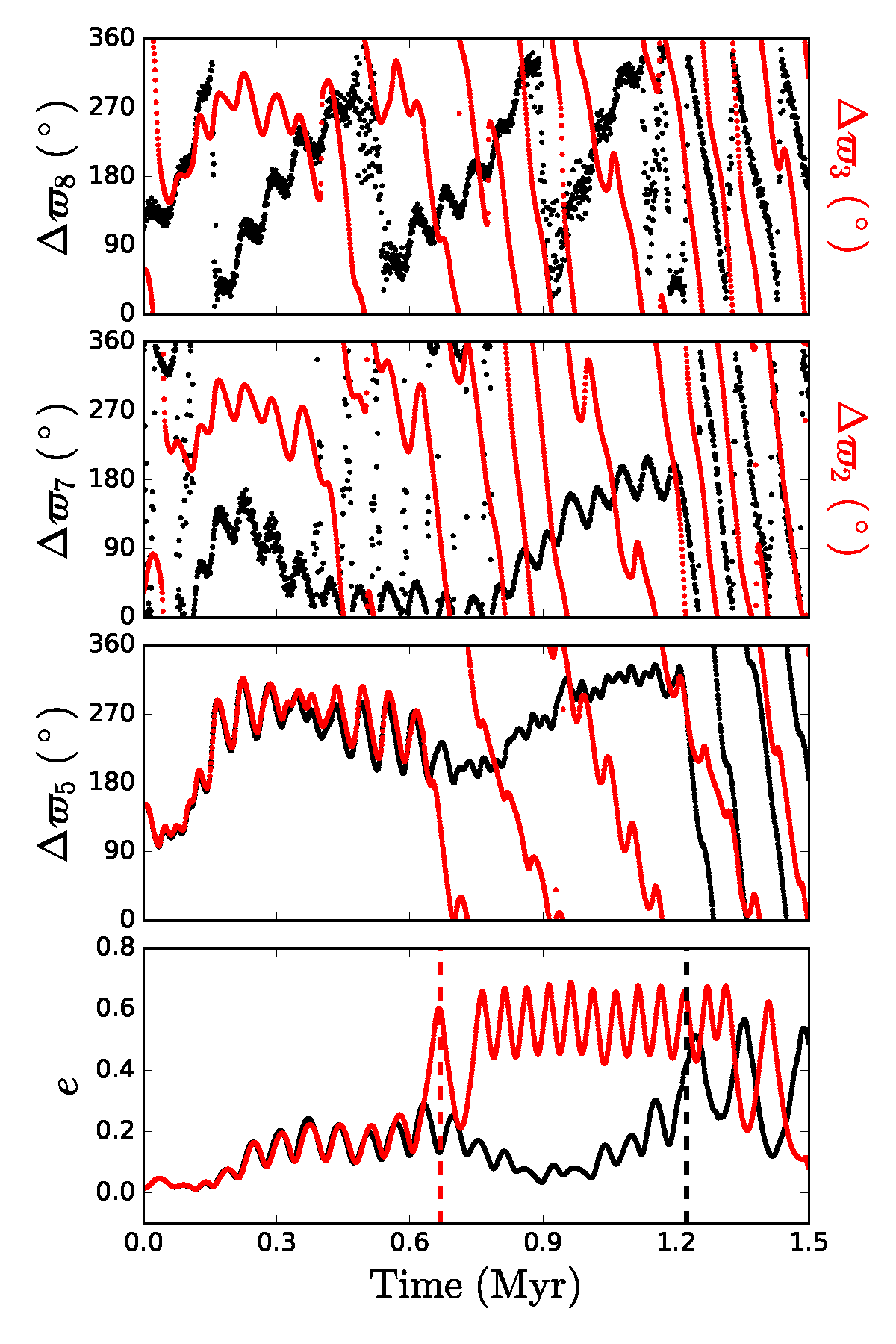}~
	\includegraphics[width=0.32\textwidth]{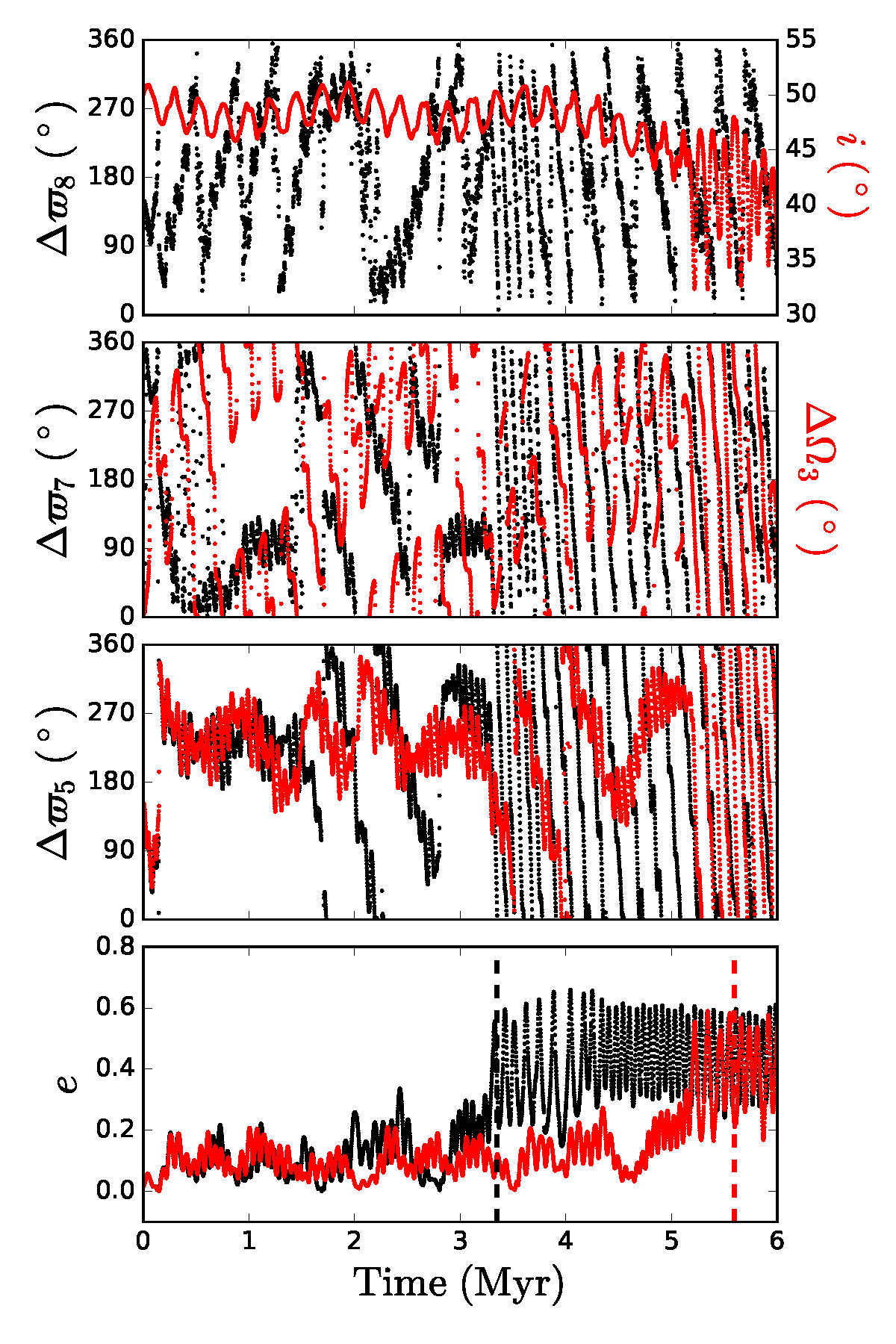}~
	\includegraphics[width=0.32\textwidth]{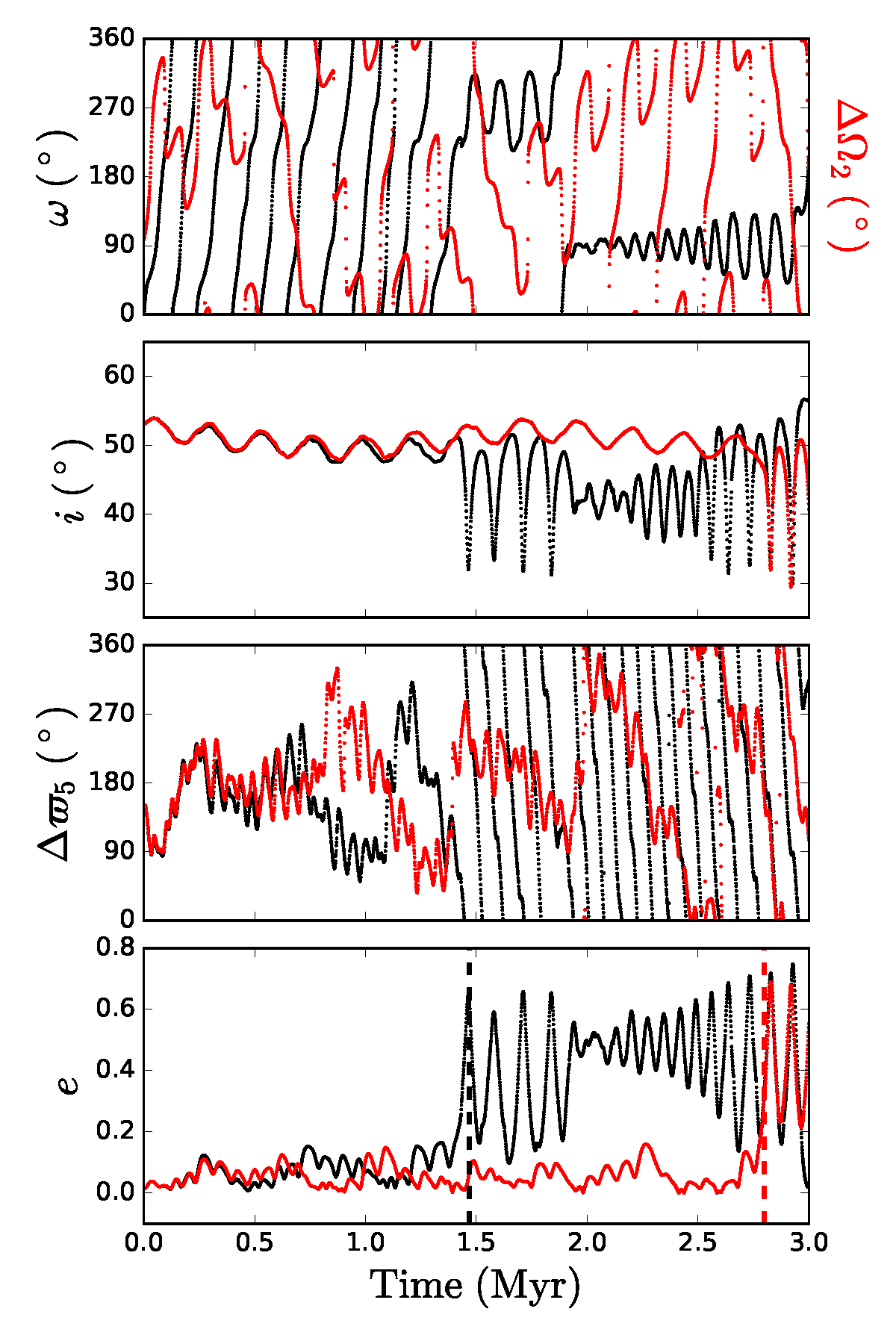}
	\caption{The orbital evolution of the fictitious Trojans marked in Fig.~\ref{fig:50D} for the model Ve2Sa (red) and Ve2Ne (black). From left to right, the three panels corresponding to the magenta, black and red stars are of the initial conditions of $(a_0,i_0)=(0.9990,48^\circ)$, $(0.9990,50^\circ)$ and $(0.9982,53^\circ)$.
	We illustrate the temporal evolution of $e$, $\Delta\varpi_5$ and $i$ for both models, $\Delta\varpi_7$, $\Delta\varpi_8$ and $\omega$ for the model Ve2Ne only and $\Delta\varpi_2$, $\Delta\varpi_3$, $\Delta\Omega_2$ and $\Delta\Omega_3$ for the model Ve2Sa only. The vertical dashed lines in the bottom plots of each panel indicate the lifespans of the orbits being in the 1:1 MMR.}
	\label{fig:50Dorb}
\end{figure*}

On a close inspection of the orbital evolution, we find the $\nu_5$ secular resonance gives a major push to the formation of the stability region. The $\nu_2$ and $\nu_3$ could be involved to some extent. In addition, the $\nu_7$ and $\nu_8$ should be responsible for the differences between the structures for two models in some ways. As examples to show these mechanisms, we display in Fig.~\ref{fig:50Dorb} the orbits with initial $(a_0, i_0)$ marked in Fig.~\ref{fig:50D} for both models.

For model Ve2Sa, $\Delta{\varpi_2}$, $\Delta{\varpi_3}$ and $\Delta{\varpi_5}$ all librate more or less between $0^\circ$--$360^\circ$ for a period of time, where $\Delta{\varpi_j}=\varpi-\varpi_j$ is the difference in the longitude of perihelion between the asteroid and some planet that is labelled with the subscript $j$. For model Ve2Ne, besides the aforementioned three critical angles, $\Delta{\varpi_7}$ and $\Delta{\varpi_8}$ show librations as well.

Given the secular critical angle, the time variation of the eccentricity can be estimated by the linear theory of secular perturbation, which gives  \citep[e.g.][]{1999ssd..book.....M,2006ChJAA...6..588L}
\begin{eqnarray}
 \begin{aligned}
\frac{de}{dt}= & C_1\sin\left(\Delta\varpi_j\right)=C_1\sin\left(\varpi-\varpi_j\right)\,, \\
C_1= & -\frac{nm_ja_je_j}{4M_\odot{a}}b^{(2)}_{3/2}\,,
\end{aligned}
\label{edot}
\end{eqnarray}
where $n$ and $m$ represent the mean motion and mass with the subscript $j$ indicating the planet. $M_\odot$ is the mass of the Sun and $b^{(2)}_{3/2}$ is a positive Laplace coefficient. Since $C_1$ is negative, ${de}/{dt}$ will be negative when $0^\circ < \Delta\varpi_j < 180^\circ$ and positive when $180^\circ < \Delta\varpi_j < 360^\circ$.

In consideration of the way in which the eccentricities evolve, we find that the $\nu_5$ secular resonance, which is characterized by a libration of $\Delta\varpi_5$, dominates the evolution of the eccentricities for both models. For model Ve2Sa, $\Delta\varpi_5$ of the most stable orbits at the blue wings librate around values $\sim180^\circ$. Under the influence of $\nu_5$, the eccentricities of these orbits almost stay unchanged according to Eq.~\eqref{edot}. For the less stable orbits in the middle area, the libration center could exceed $180^\circ$ and thus the eccentricities should undergo a secular increment. However, the participation of the $\nu_2$ and $\nu_3$ triggers the overlap of the secular resonances and the chaos will arise. Therefore, the orbits in the $i=50^\circ$ region could hardly survive a lifespan longer than 6\,Myr.

As shown in Fig.~\ref{fig:50Dorb}, for model Ve2Ne, $\Delta\varpi_7$ and $\Delta\varpi_8$ librate when the orbits remain in the Trojan clouds of the Earth. The inclusion of Uranus and Neptune sets the stage for the $\nu_7$ and $\nu_8$ secular resonances, which could further intensify the resonance overlap. Hence, the stability of the former blue areas could be reduced. Nevertheless, the perturbations from Uranus and Neptune could affect the precession of Jupiter. In some cases, the $\nu_5$ secular resonance can be maintained or even strengthened and that may also give rise to the extended stability region to the low inclinations.

The variations of the inclinations are likely under the control of the $\nu_{12}$ or $\nu_{13}$ secular resonances (see the middle and right panels of Fig.~\ref{fig:50Dorb}). The $\nu_{16}$ could be involved as well. The map for the variation of inclination reflects some structures similar to those for the maximum eccentricity. %and the stability window will disappear as all the orbits there leave the 1:1 MMR region.

From the right panel of Fig.~\ref{fig:50Dorb} we find that the Kozai mechanism would be likely to take place once the orbits leave the $\nu_5$ secular resonance. In the Ve2Ne model (black lines), the orbit leaves the trojan area due to a quick increase of eccentricity around $1.45$\,Myr, which is accompanied by a simultaneous decrease of inclination. This is the typical behaviour of orbits affected by the Kozai mechanism. The argument of perihelion $\omega$ could librate around $\pm90^\circ$ and the inclination varies in exchange of the eccentricity to keep the Delaunay variable $H_{\rm K}=\sqrt{1-e^2}\cos{i}$ constant.

\subsection{Excitation of the eccentricity and inclination}\label{subsec:excinc}
\begin{figure}[htbp]
		\centering
  \includegraphics[scale=0.27]{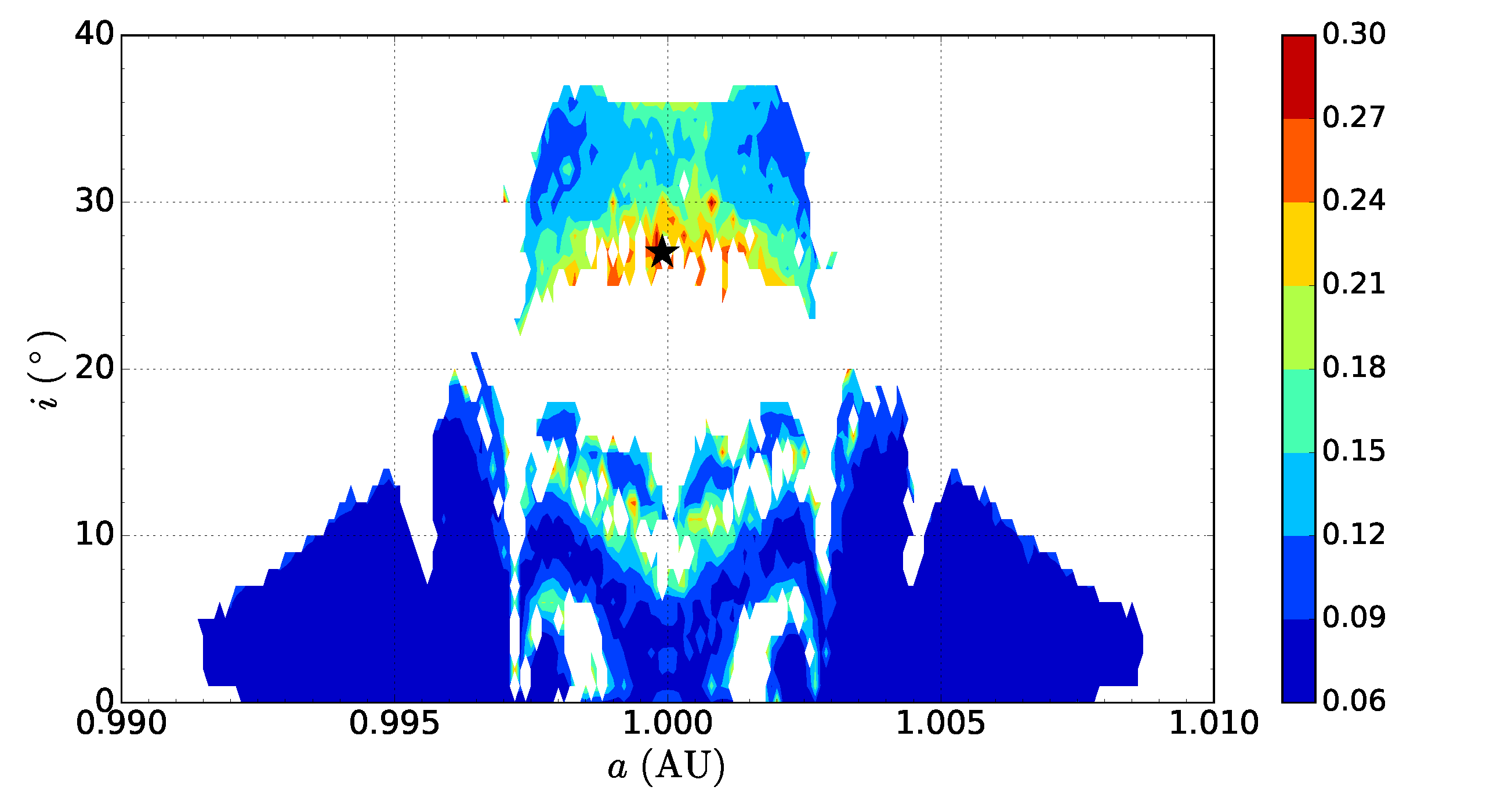}
  \caption{The maximum eccentricity (in different colours) during orbital integrations of 12\,Myr on the $(a_0,i_0)$ plane. We show only for $i_0\in[0^\circ,40^\circ]$ because there are no stable orbits with $i_0>37^\circ$. The black star stands for an orbit of $(a_0,i_0)=(0.9999,27^\circ)$ that is displayed in Fig.~\ref{fig:eccextorb}.}
  \label{fig:EMAXL4}
\end{figure}

The instability and escape of orbits in the trojan region may be caused by various mechanisms, especially secular resonances. Among those survived orbits, various dynamical mechanisms may also excite their eccentricities and inclinations. Fig.~\ref{fig:EMAXL4} shows the maximum eccentricity during orbital evolution on the $(a_0,i_0)$ plane. Owing to the longer integration time, the instability strips in Fig.~\ref{fig:EMAXL4} are much broader than the ones found in previous study (Fig.~8 in \citealt{2012A&A...541A.127D}).
%Obviously, they cannot depict the structure as detailedly as SN does. However, some clues are still provided for us to trace the resonance mechanisms under the dynamical behavior.

Fig.~\ref{fig:EMAXL4} suggests that most survived orbits obtain a maximum eccentricity smaller than 0.09 during the integration time except those in the central area. Especially, the eccentricity of the orbits at the bottom of the island around $i=30^\circ$ could reach its peak at 0.30, making it possible for them to encounter with other planets thus escape from the Earth co-orbital region in future.

\begin{figure}[htbp]
		\centering
	\includegraphics[scale=0.27]{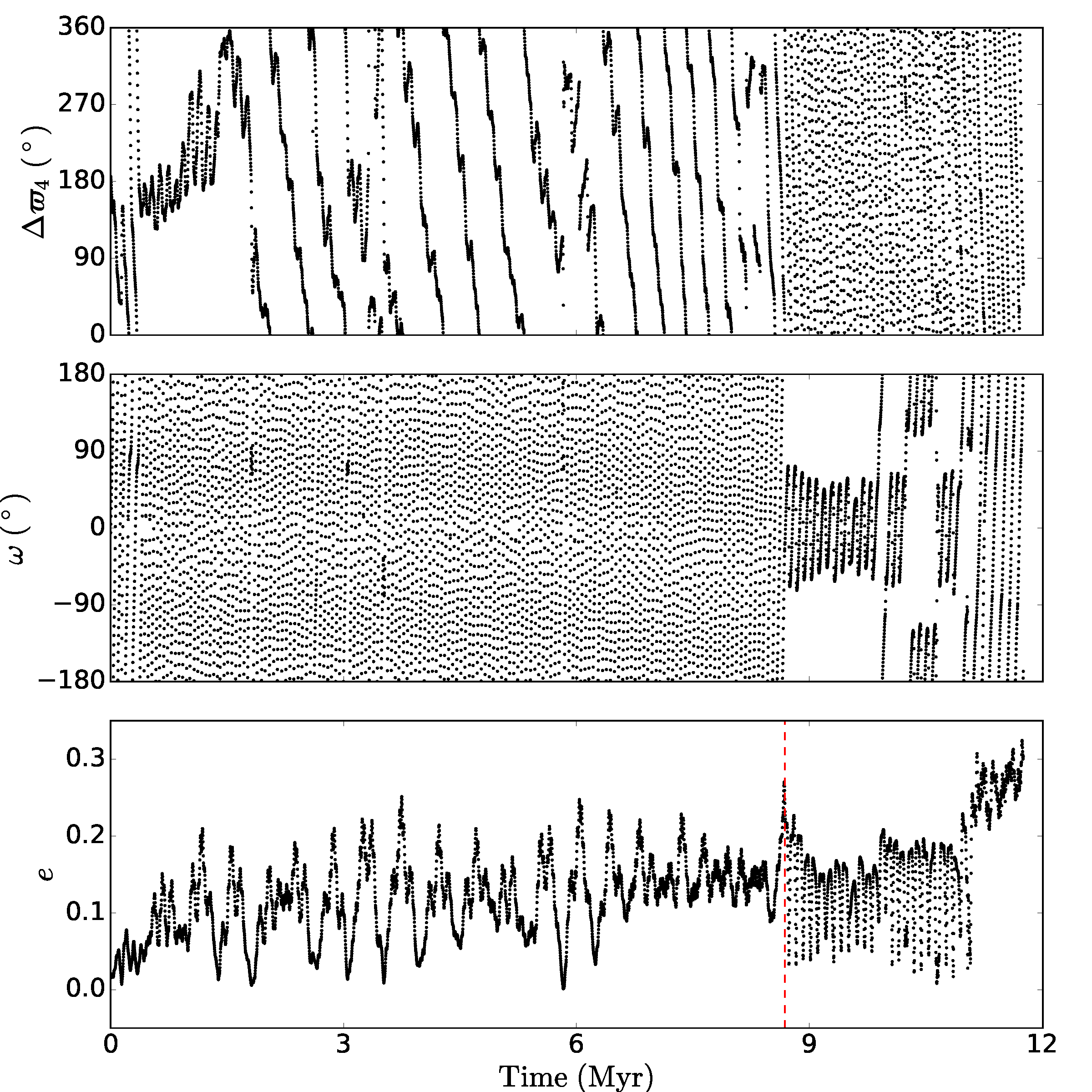}
	\caption{The orbital evolution of the fictitious Earth Trojan marked in Fig.~\ref{fig:EMAXL4} with the initial condition of $a_0=0.9999$\,AU, $i_0=27^\circ$. From top to bottom, the three panels indicate the evolutions of the apsidal differences between the Trojan and Mars $\Delta\varpi_4$, the argument of the perihelion $\omega$ and eccentricity $e$. The red vertical dashed lines in the bottom panel indicate the lifespan of the orbit.}
	\label{fig:eccextorb}
\end{figure}

To find out the involved secular resonances exciting the eccentricity, we check several possible critical angles. A preliminary inspection suggests that the $\nu_4$ secular resonance governs the motion of the orbits at the bottom of the stability island around $i=30^\circ$. The $\nu_2$ and $\nu_3$ resonances librate off and on during the whole life of these orbits. These mechanisms could act together to drive the eccentricity up with an upper limit of $\sim 0.25$. The larger eccentricity up to $\sim0.3$ can be excited by the Kozai mechanism or some higher-degree secular resonances (see Section~\ref{subsec:secres}) and as a result, the orbits will undergo close encounters with other planets and escape from the 1:1 MMR finally. As an example, we illustrate the evolution of such an orbit in Fig.~\ref{fig:eccextorb}. In fact, the survived orbits with the largest eccentricities in the stability island are of relatively large SN, which means they may leave the co-orbital region in tens of millions years and they also have a chance to be captured by the Kozai mechanism therein, just like the orbit shown in Fig.~\ref{fig:eccextorb}.

\begin{figure}[htbp]
		\centering
  \includegraphics[scale=0.27]{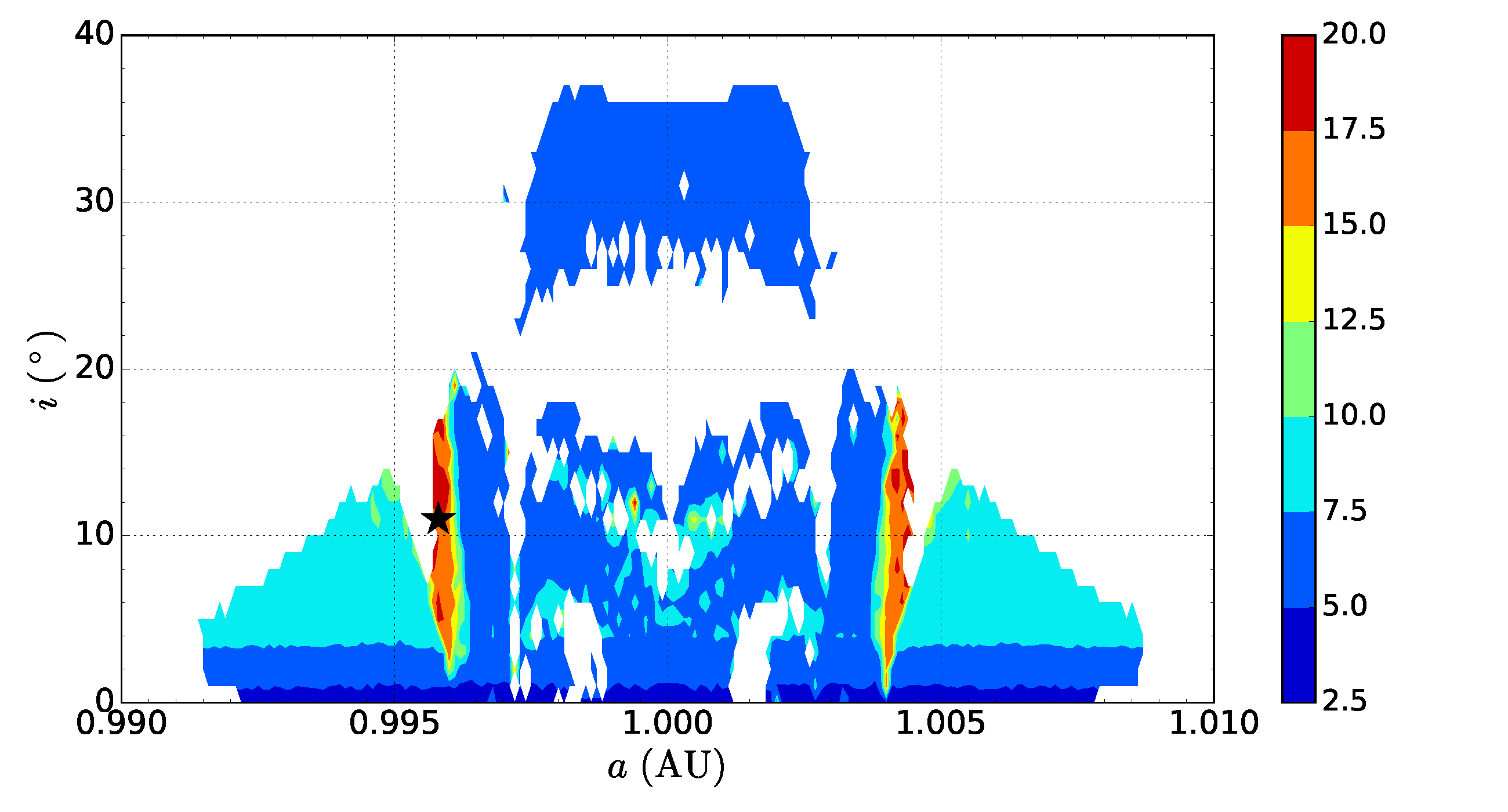}
  \caption{The same as Fig.~\ref{fig:EMAXL4} but colours indicate the variation of  inclinations. The black star stands for an orbit of $(a_0,i_0)=(0.9958,11^\circ)$ on the initial plane that is displayed in Fig.~\ref{fig:incextorb}.}
  \label{fig:DIL4}
\end{figure}

The variation of inclination during the integration time is presented in Fig.~\ref{fig:DIL4} on the $(a_0,i_0)$ plane, which clearly shows that most Trojans cannot be excited to high-inclined orbits. Two distinct vertical strips indicating the largest variations of inclinations are on the edge of the triangular instability gaps. Apparently, the orbits in the red strips are protected by some strong resonances, which could excite the inclinations as well. It is worth noting that for the orbits with $i_0<20^\circ$, the variations of the inclinations may depend on their initial inclinations. They follow a positive correlation and the inclinations of the coplanar orbits are the hardest to be excited.

As well known, the secular resonances related to the apsidal and nodal precession may excite the eccentricity and inclination respectively. Similar to the variation of the eccentricity mentioned before in Eq.\,\eqref{edot}, the evolution of the inclination can be described by \citep[e.g.][]{1999ssd..book.....M,2006ChJAA...6..588L}
\begin{eqnarray}
  \begin{aligned}
 \frac{di}{dt}= & C_2\sin\left(\Delta\Omega_j\right)=C_2\sin\left(\Omega-\Omega_j\right)\,, \\
  C_2= & \frac{nm_ja_j}{M_\odot{a}\sin{i}}b^{(1)}_{3/2}\sin\left(\frac{1}{2}i\right)\sin\left(\frac{1}{2}i_j\right)\,,
  \end{aligned}
  \label{idot}
\end{eqnarray}
where $b^{(1)}_{3/2}$ is another positive Laplace coefficient and hence $C_2$ is a positive number for prograde orbits. As a result, ${di}/{dt}$ will be positive when $0^\circ<\Delta\Omega_j<180^\circ$ and negative when $180^\circ<\Delta\Omega_j<360^\circ$.

We check carefully the possible nodal secular resonances, and find that the critical angle $\Delta\Omega_3$ of the coplanar orbits librates around $0^\circ$ with extremely small amplitudes, thus their inclinations oscillate to a well limited extent. The amplitude of $\Delta\Omega_3$ climbs towards the peak as the initial inclination increases until $i_0\sim3^\circ$, from where the angle $\Delta\Omega_3$ starts to circulate. 
The orbits in cyan at both wings of the stability region in Fig.~\ref{fig:DIL4} may be sheltered by some higher-degree secular resonances (see Section~\ref{subsec:secres}) that could drive their inclinations up at the same time.

\begin{figure}[htbp]
		\centering
	\includegraphics[scale=0.27]{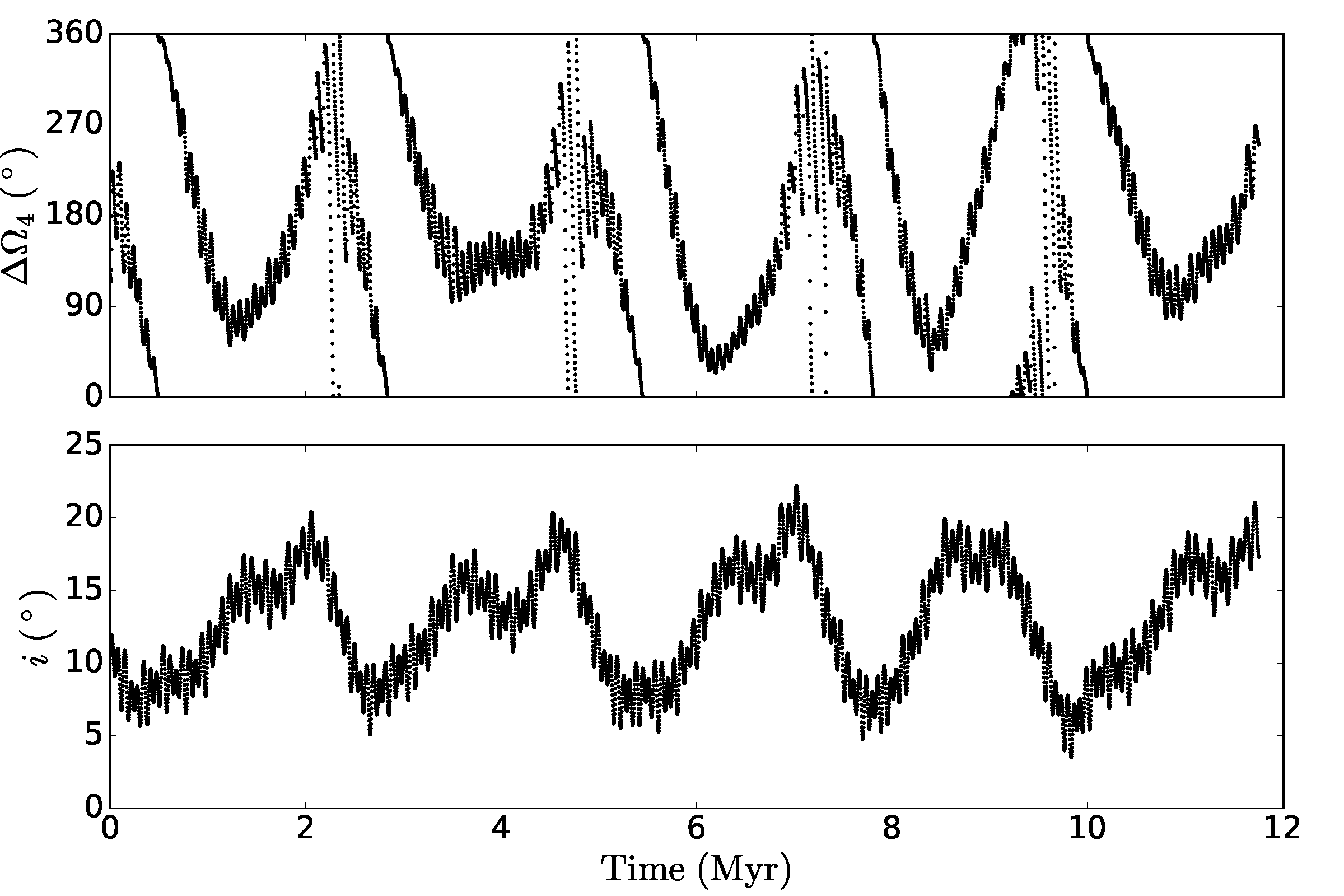}
	\caption{The evolutions of the nodal differences between the Trojan and Mars $\Delta\Omega_4$ (top panel) and inclination $i$ (bottom panel) of the fictitious Earth Trojan marked in Fig.~\ref{fig:DIL4} with the initial condition of $a_0=0.9958$\,AU, $i_0=11^\circ$. }
	\label{fig:incextorb}
\end{figure}

The $\nu_{14}$ secular resonance is dominant in exciting the inclinations to a level above $20^\circ$. As an example, Fig.~\ref{fig:incextorb} shows that the critical angle $\Delta\Omega_4$ of such an orbit could librate between $0^\circ$ and $180^\circ$ for some time and the inclination oscillates correspondingly following the rule in Eq.\,\eqref{idot}. The inclination could reach its peaks at $\sim20^\circ$ within 2\,Myr and then oscillates with a secular period of $\sim 2.7$\,Myr. Not like in the Kozai mechanism, in this case the large variation of inclination is dissociated with the eccentricity, and orbits in the corresponding region just maintaining small eccentricities as Fig.~\ref{fig:EMAXL4} shows.

Actually, some higher-degree secular resonances also play an important role in sculpting the dynamical map. With the aim of probing all possible resonances responsible for the stability of Earth Trojans, we conduct the frequency analysis in the next section.

\section{Frequency Analysis}\label{sec:fma}

Much valuable information lying behind the motions deserves to be mined as it points the way to locating the related secular and secondary resonances. In this section we follow the frequency analysis method proposed in \citet{2009MNRAS.398.1217Z,2011MNRAS.410.1849Z} to portray the resonances. For each orbit in our simulations, several most significant frequencies with the largest amplitudes of $\cos\sigma$, $e\cos\varpi$ and $i\cos\Omega$ are calculated utilizing the technique introduced in Section~\ref{subsec:naff}. Then the proper frequencies indicating the precession rates are picked out and denoted by $f_\sigma$, $g$ and $s$ respectively.

\subsection{Dynamical spectrum}\label{subsec:dyspe}

Generally, the spectra consisted of the forced frequencies, free frequencies, their harmonics and combinations are complicated. Fortunately, the proper frequencies stand out in continuous variations along some orbital parameters while the forced ones basically remain unchanged.

\begin{figure}[htbp]
	\centering
	\includegraphics[scale=0.27]{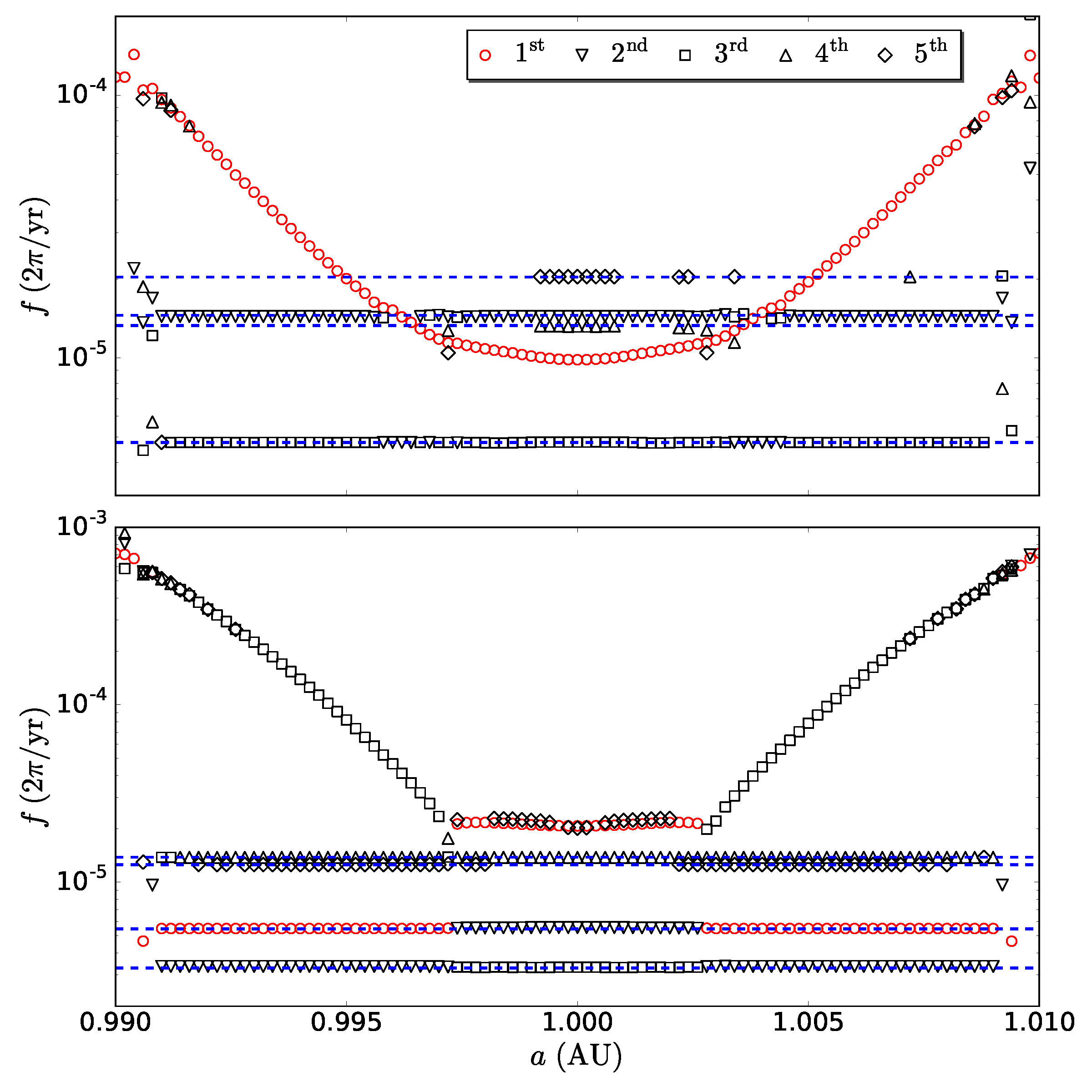}
	\caption{Dynamical spectra of $i\cos\Omega$ (top panel) and $e\cos\varpi$ (bottom panel). Both are for orbits with initial inclinations $i_0=2^\circ$. Five most significant frequencies for each orbit are plotted against their initial semi-major axes. The frequencies with the largest amplitudes are labelled by red open circles while the second to the fifth are denoted by black open inverted triangles, squares, triangles and diamonds respectively. The blue dashed lines indicate the absolute values of the fundamental frequencies identified to be corresponding to forced frequencies. From top to bottom, they represent $|s_6|$, $|s_3|$, $|s_4|$, $2g_7-s_5$, $g_4$, $g_3+s_3-s_4$, $g_2$ and $g_5$ respectively. Note that	$s_3$ and $s_4$, $\left(g_3+s_3-s_4\right)$ and $g_4$ are so close to each other that they are hardly distinguishable in the figure. Moreover, for a better vision, we just plot one third of the data.}
	\label{fig:dynspe}
\end{figure}

In our investigation, we vary the semi-major axes of the fictitious Earth Trojans with other orbital parameters fixed and record several leading frequencies for each orbit. We will obtain the so-called ``dynamical spectrum'' if we plot all these frequencies against their initial semi-major axes. The dynamical spectra of $i\cos\Omega$ and $e\cos\varpi$ for each initial inclination are calculated and two examples are shown in Fig.~\ref{fig:dynspe}. The value $i_0=2^\circ$ is selected to be illustrated here as examples arbitrarily, as well as taking into account that the stability region at low inclination is wide (see Fig.~\ref{fig:SNL4}) so that the corresponding dynamical spectra are expected to be relatively ``clearer'' for illustration. As we can see, the proper frequencies vary continuously with the semi-major axis while the forced ones derived from the perturbers hold the line. The fundamental secular frequencies in our Solar System have been calculated with different models and different methods in lots of work \citep[e.g.][]{1987A&A...181..182C,1989Nobili,1990Icar...88..266L,2006Lhotka}. In this paper we refer to the results computed by \citet{1989Nobili} and \citet{1990Icar...88..266L} (Table~\ref{tab:ffsolsys}). From the adopted fundamental frequencies, we identify the forced frequencies in the top panel as $|s_6|$, $|s_3|$, $|s_4|$ and $2g_7-s_5$.
For the bottom panel, the forced frequencies are determined to be $g_4$, $g_3+s_3-s_4$, $g_2$ and $g_5$.

\begin{table}[htbp]
		\centering
\caption{The fundamental secular frequencies in the Solar System except for Mercury. The periods are given in years and the frequencies are given in $10^{-7}\,2\pi\,{\rm yr}^{-1}$. The period values of the inner planets are taken from \citet{1990Icar...88..266L} while those of the outer planets are taken from \citet{1989Nobili}. The frequencies are computed from the corresponding periods. }
\begin{tabular}{crr|crr}
	\hline
  & Period~~~ & Freq.~ &  & Period~~~ & Freq.~ \\
  \hline
$g_2$ & 173,821.61 & 57.53 & $s_2$ & $-183,063.27$ & $-54.63$\\
$g_3$ & 74,634.47 & 133.99 & $s_3$ & $-68,749.02$ & $-145.46$ \\
$g_4$ & 72,339.20 & 138.24 & $s_4$ & $-73,021.30$ & $-136.95$ \\
$g_5$ & 304,400.48 & 32.85 & $s_5$ & $-129,550,000.$ & $-0.08$ \\
$g_6$ & 45,883.37 & 217.94 & $s_6$ & $-49,193.46$ & $-203.28$ \\
$g_7$ & 419,858.29 & 23.82 & $s_7$ & $-433,059.42$ & $-23.09$ \\
$g_8$ & 1,926,991.9 & 5.19 & $s_8$ & $-1,871,442.70$ & $-5.34$ \\
  \hline
\end{tabular}
\label{tab:ffsolsys}
\end{table}

Retrograde precession implies negative frequency values. The fundamental frequencies $s_j$ corresponding to the nodal precession rates of the planets are all negative. However, all frequencies we obtain by the numerical analysis should be positive in principle, so we have to monitor the variations of the longitude of perihelion and the longitude of ascending node to determine the sign of $g$ and $s$ for each fictitious Trojan.

The proper frequencies are not necessarily the ones with the largest amplitudes (red open circles in Fig.~\ref{fig:dynspe}). The motions could be governed by some forced frequencies just like the orbits on both sides in the bottom panel, which are dominated by the nodal precession of Venus with $g_2=5.753\times10^{-6}~2\pi\,{\rm yr}^{-1}$.

A linear secular resonance occurs every time the proper frequency $g$ or $s$ comes across the fundamental frequencies. High-degree resonances involving the combinations of secular frequencies can be determined in a similar way. However, the dynamical spectra could not always be as ``clear'' as the examples in Fig.~\ref{fig:dynspe}. Actually, orbital chaos could make the dynamical spectra so complicated that we can only pick out the proper frequencies accurately in a limited area. Another noticeable feature in Fig.~\ref{fig:dynspe} that will be discussed in the following part is that the proper frequencies seem to be divided into three regimes and we will see they are actually separated by the boundary between the tadpole and horseshoe orbits. This feature is more obvious for $e\cos\varpi$ and $\cos\sigma$ but a close inspection proves it is valid for all three variables.

\subsection{Tadpole and horseshoe orbits}\label{subsec:tadhorse}

For the planar circular restricted three-body problem, the test particles with small eccentricities are presumed to move in tadpole orbits if their orbits satisfy the equation \citep{1999ssd..book.....M}
\begin{equation}
  \delta{r}\leq\left(\frac{8}{3}\mu\right)^{1/2}a\,,
\end{equation}
where $\delta{r}$ is the radial separation of the particles from the secondary mass, whose mass fraction and semi-major axis are denoted by $\mu$ and $a$. For the Sun-Earth system, $\mu\approx3\times10^{-6}$ and $a=1$\,AU, thus $\delta{r}\approx 0.00283$\,AU. Recall the phase portrait (see Fig.\,2 in \citealt{2012A&A...541A.127D}) derived from the symplectic mapping method and we will get the same criterion considering $\delta{r}\approx\delta{a}$ for near-circular orbits.
\citet{1974AJ.....79..404W} has demonstrated that the tadpole orbits could be found within 0.00285\,AU away from the Earth while the horseshoe regime extends to $1\pm0.0080$\,AU. A larger horseshoe orbit could lose its stability soon due to the close approach to the Earth near the turning points.

In our study we could check the variation of the critical angle $\sigma=\lambda-\lambda_3$ of the 1:1 MMR for each stable orbit from the output of the simulation to determine whether or not it is in the tadpole regime. Actually another way taking advantage of the frequency analysis is preferred in practice to define the separatrices. We numerically calculate the spectrum of the variable $\cos\lambda$ for each orbit and record the dominant frequency of which the amplitude is the largest. For tadpole orbits the dominant frequency should exactly equal to that of the mean motion of the Earth while the dominant frequency for horseshoe orbits could differ from the orbital frequency of the Earth owing to their elongated trajectories, which are farther away from the Earth. We adopt the above criterion and locate the critical semi-major axis for different inclinations. A numerical fit of the separatrices are
\begin{eqnarray}
  \begin{aligned}
    a_{\rm low}= & 0.9972-1.738\times10^{-6}\,i+1.931\times10^{-7}\,i^2 \\
                 & -8.449\times10^{-10}\,i^3\,,\\
    a_{\rm up}=  & 1.0030-8.621\times10^{-6}i +1.877\times10^{-7}\,i^2 \\
                 & -3.091\times10^{-9}\,i^3 \,,
  \end{aligned} \label{eq:alowup}
\end{eqnarray}
where $a_{\rm low}$ and $a_{\rm up}$ represent the lower and upper limit of the semi-major axis for tadpole orbits.

The separatrices between the tadpole and horseshoe orbits are shown in Fig.~\ref{fig:SNL4} utilizing Eq.\,\eqref{eq:alowup}. Obviously, the separatrices perfectly fit the instability rifts around $1\pm 0.0028$\,AU in the dynamical map. As we know, motions near the separatrix in the perturbed system are so chaotic that the instabilities will arise there. Moreover, recall Fig.~\ref{fig:dynspe} and we find the proper frequencies undergo piecewise changes along semi-major axes. The variation trends are nearly symmetrical in region ``L'' and ``R'' while they appear greatly different for tadpole orbits in region ``C''. In the stability island around $30^\circ$ where the high-inclined orbits could survive the integration time ($\sim12$\,Myr), all fictitious Trojans are in tadpole orbits.

\subsection{Empirical formulae}\label{subsec:empfom}

Expressions of the proper frequencies $g$ and $s$ on the plane $(a_0,i_0)$ are necessary for constructing the resonance maps. We pick out the orbits whose proper frequencies can be accurately determined and obtain the empirical formulae by numerical fitting.
In order to reduce the round-off error, we normalize $a_0$ and $i_0$ of the selected orbits as $x=(a_0-\bar{a}_0)/\sigma_{a_0}$ and $y=(i_0-\bar{i}_0)/\sigma_{i_0}$, where $\bar{a}_0$, $\bar{i}_0$ and $\sigma_{a_0}$, $\sigma_{i_0}$ represent the mean value and standard deviation of the initial elements respectively (Table~\ref{tab:mstd}).

\begin{table}
		\centering
	\caption{The mean values and standard deviations of $a_0$ (in AU) and $i_0$ (in degree) of the orbits selected for the numerical fitting for $g$ and $s$ respectively in region L, C and R. $\bar{a}_0$ and $\bar{i}_0$ stand for mean values while $\sigma_{a_0}$ and $\sigma_{i_0}$ stand for standard deviations. We adopt the double-precision floating-point format in our calculations but we only show 6 significant figures in this table for lack of space. }
	\begin{tabular}{cc|cccc}
    \hline
    {} & {} & $\bar{a}_0$ & $\sigma_{a_0}$ & $\bar{i}_0$ & $\sigma_{i_0}$ \\
    \hline
	{ } & L & 0.995040 & 0.00173011 & 9.29600 & 8.40849 \\
    {g} & C & 1.00002 & 0.00166711 & 17.0435 & 12.1155 \\
    { } & R & 1.00508 & 0.00176528 & 9.24342 & 8.35572 \\
    \hline
    { } & L & 0.995299 & 0.00166176 & 13.8992 & 12.1669 \\
    {s} & C & 1.00002 & 0.00159960 & 26.5344 & 15.9284 \\
    { } & R & 1.00479 & 0.00169481 & 13.5209 & 12.0755 \\
    \hline
	\end{tabular}
	\label{tab:mstd}
\end{table}

We adopt the quintic polynomial as follows,
\begin{equation}
  f(x,y)=\sum_{m=0}^{5}\sum_{n=0}^{5-m}\,p_{mn}x^my^n\,.
\end{equation}
The coefficients $p_{mn}$ of the best fits for three different regions are listed in Table~\ref{tab:fitpara}. We fit the proper frequencies in region ``L'' and ``R'' separately to improve precision although they seem symmetrical about $a_0\approx 1$\,AU. Only the proper frequencies of the regular orbits can be determined accurately and used for the numerical fitting. In the white areas between the stability regions, where the chaos arise, the empirical formulae are less reliable because the frequency spectra of those orbits are disordered and the proper frequencies can only been gained by interpolating. In the marginal area far away from the stability regions, the proper frequencies derived from the empirical formulae are suggested for reference only due to a complete extrapolation.

\begin{table*}
\caption{The coefficients of the best fits for the proper frequencies $g$ and $s$ (in $2\pi\,{\rm yr}^{-1}$) in region L, C and R on the $(a_0,i_0)$ plane. The quintic polynomials are adopted to numerically fit $g$ and $s$ and to save space here only the coefficients  $p_{mn}$ of the term $x^{m}y^{n}$ for $m+n\leq 3$ are listed. All coefficients have been multiplied by a factor of $10^{8}$.}
	\centering
\begin{tabular}{cc|rrrrrrrrrr}
\hline
{} & {} & $p_{00}$~~ & $p_{10}$~~ & $p_{01}$~~ & $p_{20}$~~ & $p_{11}$~~ & $p_{02}$~~ & $p_{30}$~~ & $p_{21}$~~  & $p_{12}$~~ & $p_{03}$~~ \\
\hline
 {} & L & $6301$ & $-4581$ & $-3004$ & $399.6$ & $5884$ & $-1135$ & $1219$ & $-6272$ & $3297$ & $-53.53$ \\
 {g} & C & $1562$ & $-0.7153$ & $-540.8$ & $142.2$ & $-0.6774$ & $24.12$ &  $0.2887$ & $-43.03$ & $-0.5720$ & $60.28$  \\
 {} & R & $6440$ & $4656$ & $-3082$ & $327.5$ & $-6140$ & $-1223$ & $-1335$ & $-6919$ & $-3800$ & $-186.2$ \\
 \hline
 {} & L & $-1568$ & $724.8$ & $386.0$ & $-295.0$ & $-312.4$ & $78.70$ & $72.39$ & $115.2$ & $-2.647$ & $-65.88$ \\
 {s} & C & $-647.2$ & $0.8774$ & $216.6$ & $-56.59$ & $-0.3256$ & $-48.78$ & $0.008195$ & $18.75$ & $-0.2804$ & $-3.840$ \\
 {} & R & $-491.8$ & $-1162$ & $-253.1$ & $1081$ & $1057$ & $335.8$ & $-638.0$ & $-969.7$ & $-665.6$ & $-74.52$ \\
  \hline
\end{tabular}
\label{tab:fitpara}
\end{table*}

\subsection{Secular resonances}\label{subsec:secres}

With the help of the empirical formulae, the secular resonances can be determined by solving the equation:
\begin{equation} \label{eq:secres}
  pg+qs+\sum_{j=2}^{8}(p_jg_j+q_js_j)=0\,,
\end{equation}
where $p$, $q$, $p_j$, $q_j$ are integers. The d'Alembert rules require $p+q+\sum_{j=2}^{8}(p_j+q_j)=0$ and $(q+\sum_{j=2}^8q_j)$ must be even. $|p|+|q|+\sum_{j=2}^{8}\left(|p_j|+|q_j|\right)$ is defined as the degree of the secular resonance.

\begin{figure}[htbp]
		\centering
	\includegraphics[scale=0.27]{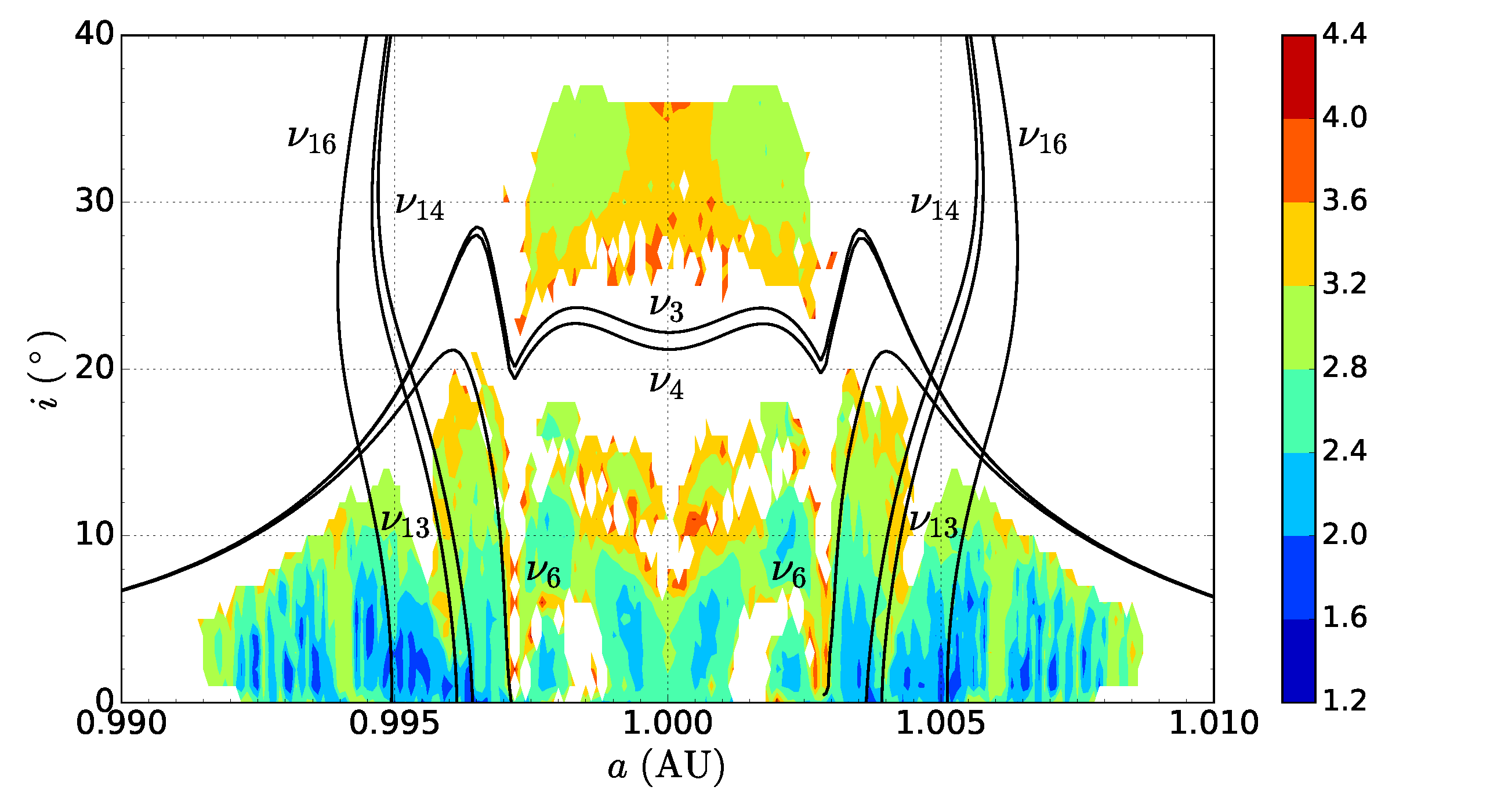}
	\caption{The locations of main linear secular resonances for Trojans around the $L_4$ point on the $(a_0,i_0)$ plane. The resonances are labelled along the curves.}
	\label{fig:secord2}
\end{figure}

A complete search of the combinations of $p$, $p_j$, $q$ and $q_j$ for different degrees will identify the secular resonances responsible for the structures in the dynamical maps. Fig.~\ref{fig:secord2} presents the main linear secular resonances on the $(a_0,i_0)$ plane. Apparently, the orbits in the $15^\circ\sim24^\circ$ instability gap are likely governed by the  apsidal secular resonances $\nu_3$ and $\nu_4$. Examinations of the critical angles reveal that the $\nu_4$ secular resonance has a stronger influence on the motions there.
The nodal secular resonances $\nu_{13}$ and $\nu_{14}$ locate exactly where the inclinations are excited. As mentioned in Section~\ref{subsec:excinc}, the $\nu_{13}$ could control the variations of the inclination depending on the initial values while the $\nu_{14}$ could drive the inclinations up to $\sim20^\circ$. The $\nu_6$ resonance corresponds to an elongated instability strip where the eccentricity can be elevated. To a lesser extent, the $\nu_{16}$ contributes to the formation of the vertical strips of stability in the dynamical map.

These linear secular resonances have strong influences on the orbital evolution of Earth Trojans and they form the major structures in the dynamical map in a relatively short time. Still, there are rich fine structures in the dynamical map, arising from higher-order secular resonances. As examples, we depict below some major ones.

\begin{figure*}[htbp]
	\centering
	\includegraphics[scale=0.28]{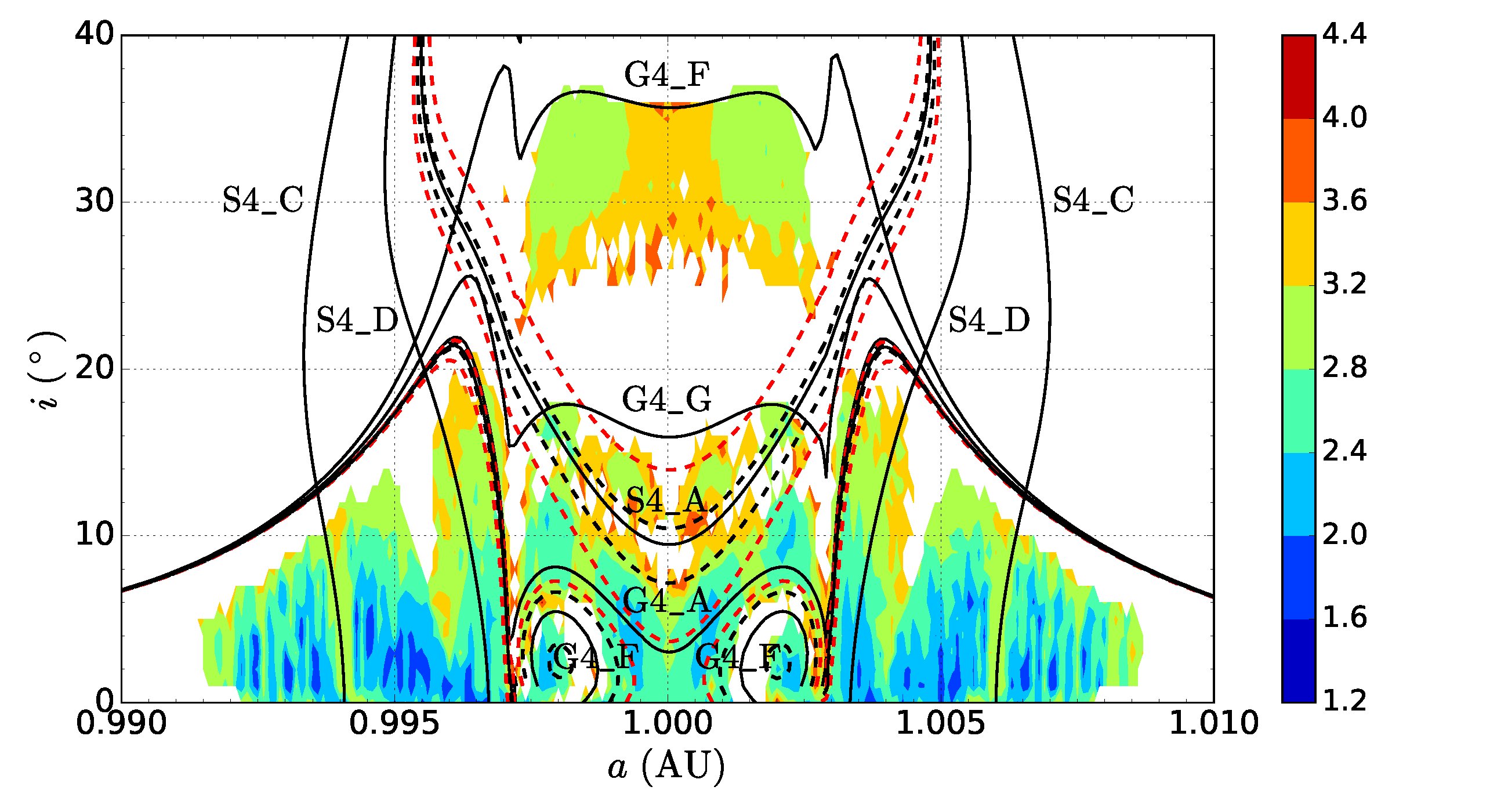}~~~
	\includegraphics[scale=0.28]{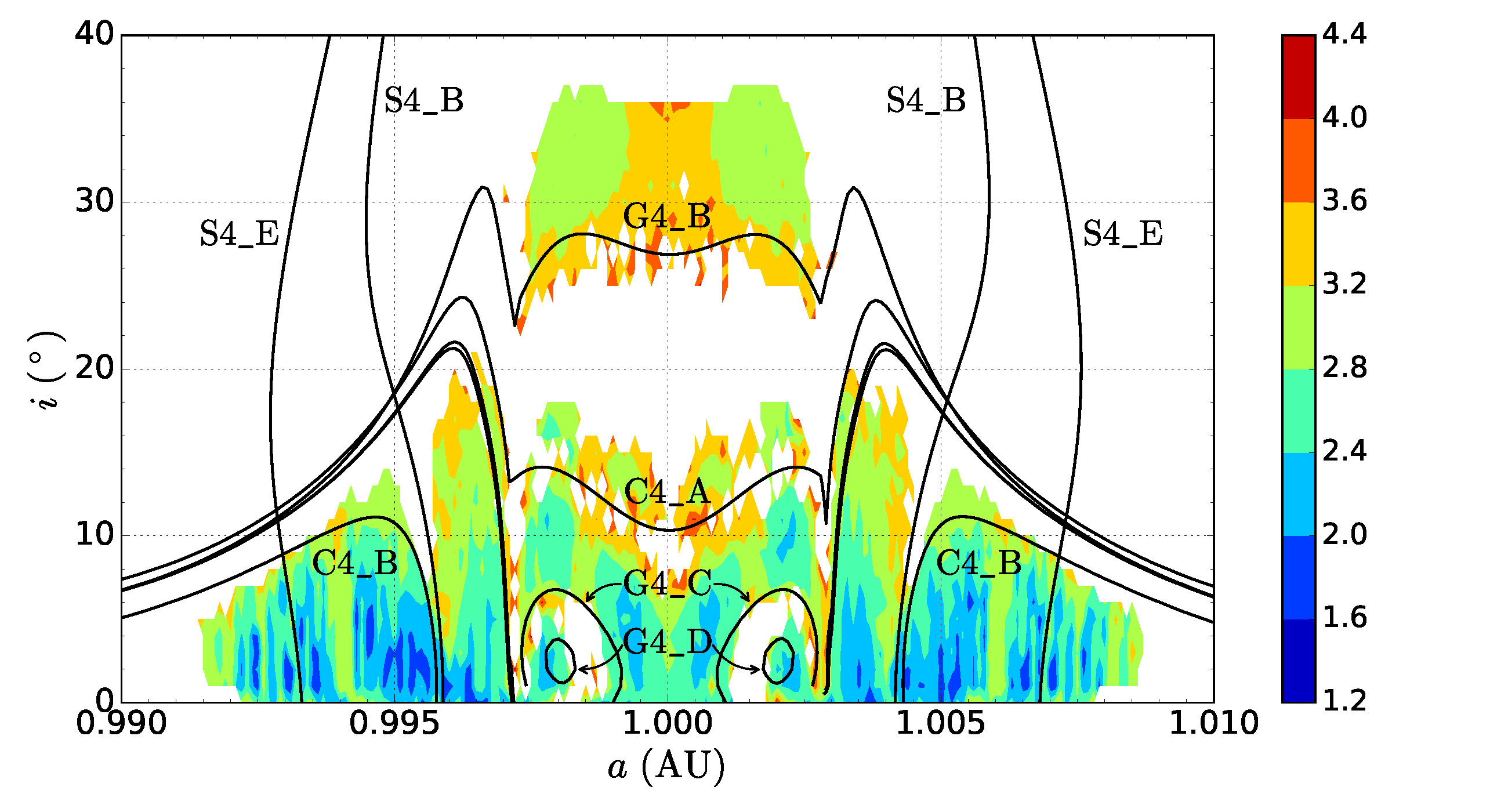}
	\caption{The locations of the fourth-degree secular resonances for Trojans around the $L_4$ point on the $(a_0,i_0)$ plane. The resonances are labelled (see text for the meaning) along the curves. They are plotted in two panels for a clear vision. In consideration of the frequency drift of the inner planets, the dashed lines in the left panel delimit the possible locations of S4\_A and G4\_F (see Section~\ref{subsec:fdrift}). Black and red dashed lines represent the results derived from the data taken from \citet{1990Icar...88..266L} and from our integration respectively.}
	\label{fig:secord4}
\end{figure*}

The main fourth-degree secular resonances are shown in Fig.~\ref{fig:secord4} and the meaning of the labels along the curves are listed as follows,
\begin{equation}
  \label{eqn:ord4}
  \begin{aligned}%{array}{llll}
    {\rm G4\_A}:  & ~ g-g_2-s_2+s_6=0, & ~
    {\rm G4\_B}:  & ~ g-g_2-s_3+s_6=0, \\
    {\rm G4\_C}:  & ~ g+g_2-2g_3=0, &  ~
    {\rm G4\_D}:  & ~ g-s_2-g_3+s_4=0, \\
    {\rm G4\_E}:  & ~ g-g_3+s_3-s_6=0, & ~
    {\rm G4\_F}:  & ~ g-g_3+g_4-g_6=0, \\
    {\rm G4\_G}:  & ~ g-g_4-s_5+s_7=0, & ~
    {\rm S4\_A}:  & ~ s+s_2-s_3-s_5=0, \\
    {\rm S4\_B}:  & ~ s-2s_3+s_4=0, & ~
    {\rm S4\_C}:  & ~ s-s_3-s_4+s_5=0, \\
    {\rm S4\_D}:  & ~ s+s_3-2s_4=0, & ~
    {\rm S4\_E}:  & ~ s-g_5+g_6-s_6=0, \\
    {\rm C4\_A}:  & ~ g-s-g_4+s_4=0, & ~
    {\rm C4\_B}:  & ~ g+2s-g_6=0.
  \end{aligned}
\end{equation}
The resonances with $q=0$ are classified as ``G'' type while those with $p=0$ are classified as ``S'' type. They represent the secular resonances only involving the precession rates of the perihelion longitudes ($g$) and ascending nodes ($s$) of Earth Trojans respectively. The others are denoted by ``C'' type in the name of ``combined''.

As we can see in Fig.~\ref{fig:secord4}, the recognized secular resonances could match the fine structures of the dynamical maps well. The resonance overlap sets the border of the stability region for orbits with moderate inclinations. In the central areas, secular resonances such as G4\_F can help block the way of the stability region towards higher inclinations. The orbits in the central instability areas are mainly subjected to the ``C-type'' and ``G-type'' secular resonances although a few ``S-type'' resonances such as S4\_A may be involved as well. The vertical strips in the region ``L'' and ``R'' of the dynamical map indicating the most stable orbits should be related to the ``S-type'' secular resonances. An obvious example is the green gaps around $1\pm 0.006$\,AU splitting two blocks of the most stable orbits. The secular resonance S4\_C involving the nodal precession of the Earth, Mars and Jupiter could destabilize the orbits in these gaps slightly. To the contrary, there may exist some other resonances improving the stability. The competition between the protection and destruction provided by the secular resonances result in the fragmentation in shape of the dynamical map.

\begin{figure*}[htbp]
		\centering
\includegraphics[scale=0.28]{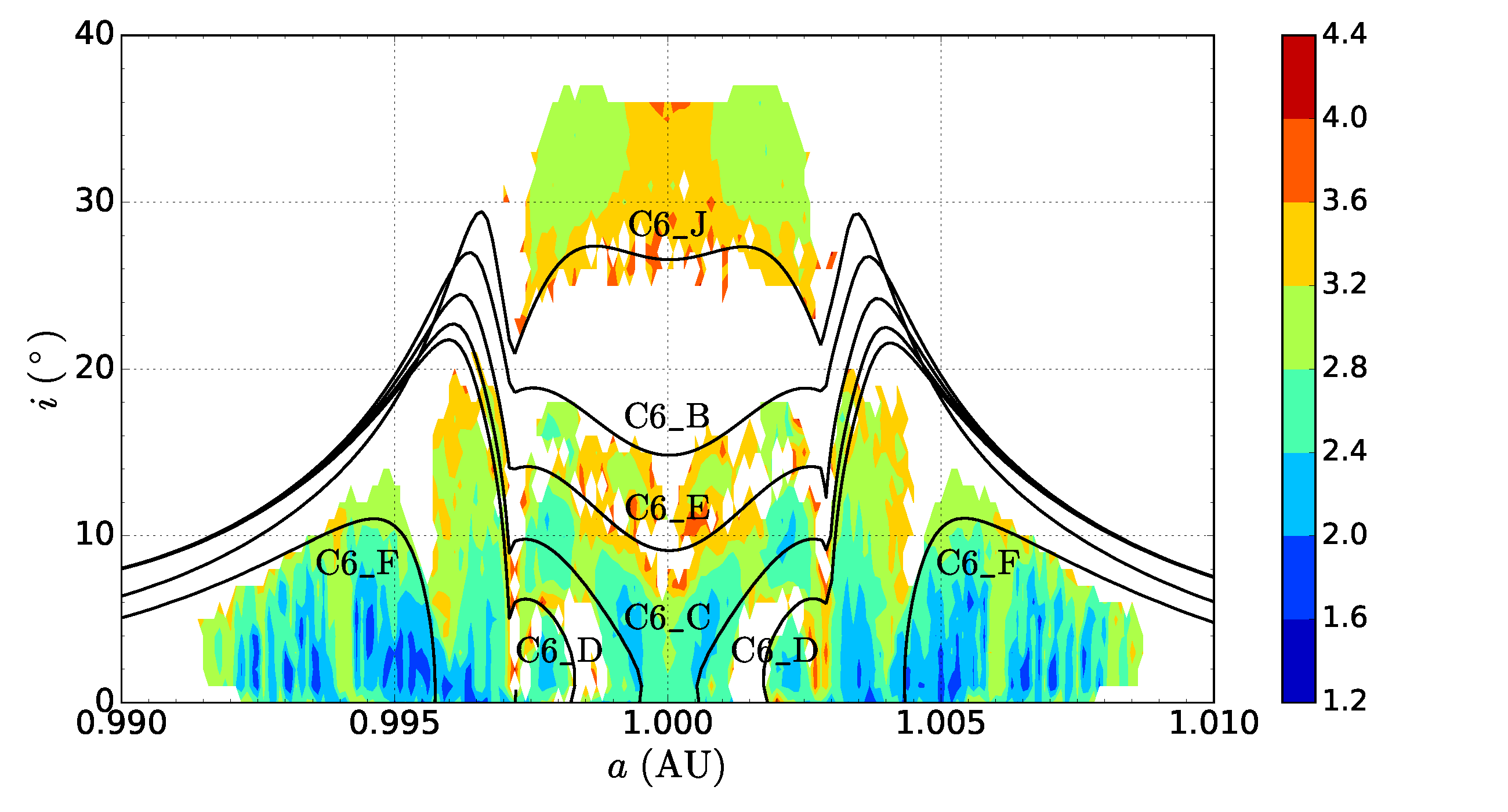}~~~
\includegraphics[scale=0.28]{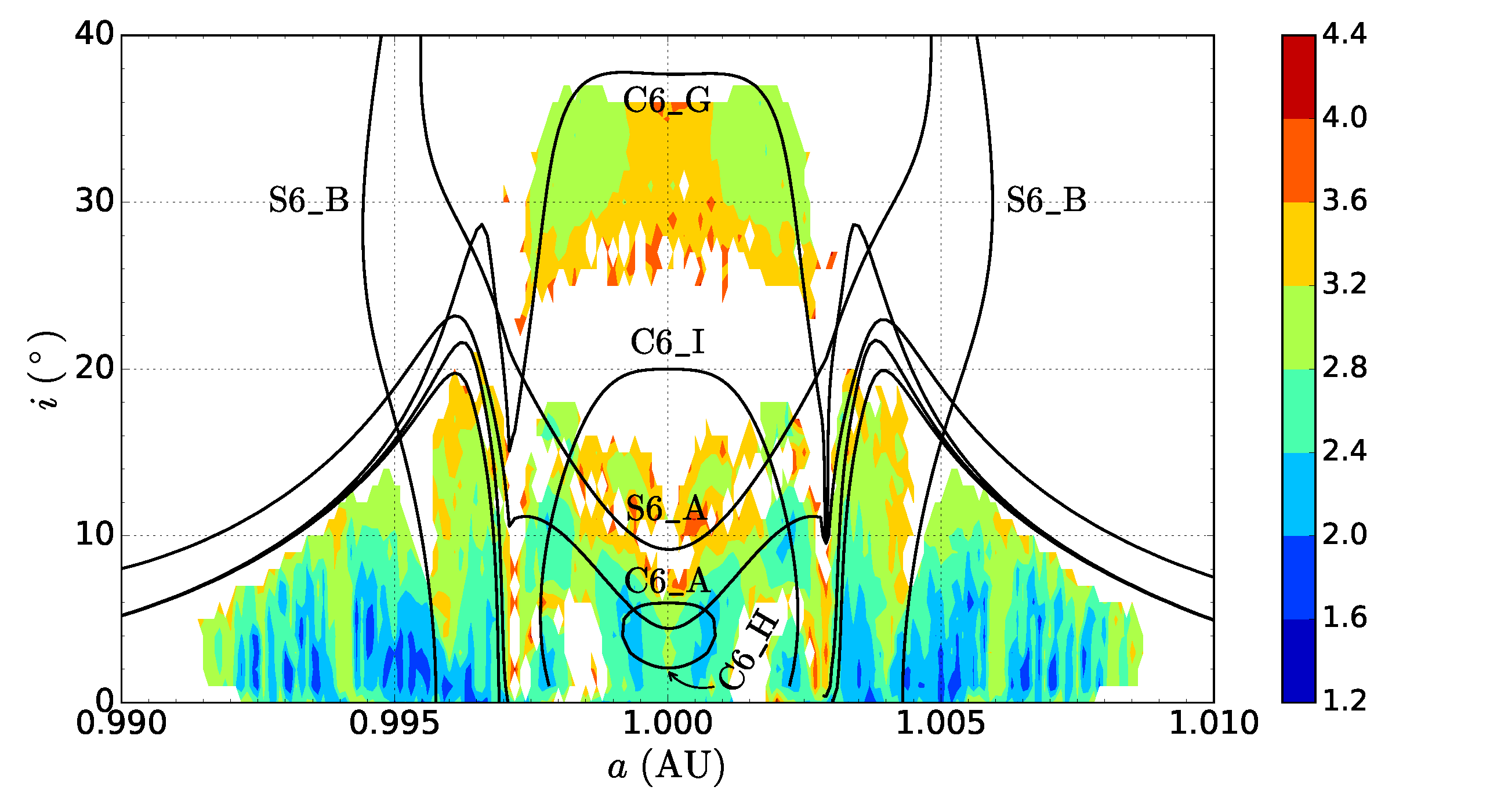}
	\caption{The same as Fig.~\ref{fig:secord4} but for sixth-degree secular resonances. }
	\label{fig:secord6}
\end{figure*}

In Fig.~\ref{fig:secord6} we illustrate the main sixth-degree secular resonances and they are explained as follows,
\begin{eqnarray}
    \label{eqn:ord6}
  \begin{aligned}%{ll}
    {\rm S6\_A}:  &  3s-2s_4-s_5=0,  &
    {\rm S6\_B}:  &  3s-2s_4-s_6=0, \\
    {\rm C6\_A}:  &  g-2s-g_2+s_4+s_6=0, &
    {\rm C6\_B}:  &  g-2s-g_2+2s_4=0, \\
    {\rm C6\_C}:  &  g-2s-g_3+2s_4=0, &
    {\rm C6\_D}:  &  g-2s+2s_3-g_4=0, \\
    {\rm C6\_E}:  &  g-2s+s_4-g_5+s_6=0, &
    {\rm C6\_F}:  &  g+2s-g_2-g_3-g_5=0, \\
    {\rm C6\_G}:  &  g+2s-s_2-g_6-s_6=0, &
    {\rm C6\_H}:  &  g+2s-g_5-s_5-s_7=0, \\
    {\rm C6\_I}:  &  g+2s-g_8-2s_8=0, &
    {\rm C6\_J}:  &  2g+s-g_2-g_3-s_7=0.
  \end{aligned}
\end{eqnarray}

It reflects that each fine structure could be shaped by multiple resonance mechanisms and actually, we just show in Figs.~\ref{fig:secord4} and \ref{fig:secord6} some representatives of the resonance families whose strength are relatively great. The secular resonances in the same family share the similar locations and they work together to control the motions of Earth Trojans, such as G4\_F and C6\_G, G4\_B and C6\_J, C4\_A and C6\_E. What also needs to be emphasized is that each resonance has a width in which the orbits can be influenced effectively. The resonance width varies greatly for different resonances so that it is not always possible to distinguish the dominant resonances in fact.

The inclusion of Uranus and Neptune completes the secular resonance net displayed above. On one hand, Uranus and Neptune directly affect the Trojans as they are involved in many secular resonances although most of them are not plotted in Figs.~\ref{fig:secord4} and \ref{fig:secord6}; on the other hand, the perturbations from Uranus and Neptune act on the motion of Jupiter and Saturn, causing a modification to the dynamical behavior of the Trojans.

In this paper we just focus on the secular resonances and no secondary resonance has been shown. Some resonances contributing to the motions may not be included in the map either because they locate outside the $(a_0,i_0)$ area. However, they can still influence the orbits within the resonance width in varying degrees, just like the $\nu_5$ resonance. According to Fig.~\ref{fig:dynspe}, the $\nu_5$ secular resonance has a great influence on some Trojans although it resides outside the $(a_0,i_0)$ area (Fig.~\ref{fig:secord2}). On the contrary, some resonances also have a chance not to occur near the locations derived from the empirical formulae if the orbits are affected more strongly by other mechanisms.

The typical lifetime of Earth Trojans trapped in a secular resonance is 1\,Myr while the typical Lyapunov time could be 1\,kyr \citep{2002MNRAS.334..241B}. The integration time in our simulations ($\sim 12$\,Myr) is long enough for the Trojans govern by some secular resonance causing instabilities to escape from the 1:1 MMR region. Moreover, considering the frequency drift of the inner planets causing the shift of the secular resonances (see next subsection), the short typical lifetime of 1\,Myr could facilitate the expansion of the instabilities.

\subsection{The frequency drift of the inner planets}\label{subsec:fdrift}

The chaotic motion with a maximum Lyapunov exponent of $\sim1/5~{\rm Myr}^{-1}$ was detected in the inner Solar System by \citet{1989Natur.338..237L} and they found it is related to the transition from libration to circulation of the critical angle of the secular resonance $2\left(g_4-g_3\right)-\left(s_4-s_3\right)=0$ \citep{1990Icar...88..266L}. As a consequence, the secular frequencies of the inner planets keep changing to make the related secular resonances sweep across a large area of the phase space over a long time span.

The mean values and the amplitudes of the variations of the fundamental secular frequencies for Venus, the Earth and Mars are listed in Table~\ref{tab:fredrift} \citep{1990Icar...88..266L}. They were obtained from the numerical computations over 200\,Myr, therefore the values of $\Delta\nu$ here just set a lower limit of the frequency drift.  %and for a longer time, they may have a great increment.
Actually, we have integrated the Solar System to 1\,Gyr and we find the frequency drift could be several times larger than the values obtained by \citet{1990Icar...88..266L}. However, we still refer to the data taken from \citet{1990Icar...88..266L} to avoid the conflict with the adopted fundamental frequencies.

\begin{table}[htbp]
	\caption{The mean values ($\bar{\nu}$) and the amplitudes of the variations ($\Delta\nu$) of the fundamental secular frequencies in the inner Solar System except for Mercury over 200\,Myr. The frequencies are given in $10^{-7}\,2\pi\,{\rm yr}^{-1}$. The data are taken from \citet{1990Icar...88..266L}.}
\centering
	\begin{tabular}{crr|crr}
\hline
   {} &  $\bar{\nu}$~~~ & $\Delta\nu$~~~ & {}  & $\bar{\nu}$~~~ & $\Delta\nu$~~~ \\
\hline
  $g_2$ & $57.52$  & $0.1003$ & $s_2$ & $-54.01$  & $1.7747$ \\
  $g_3$ & $133.49$ & $1.3117$ & $s_3$ & $-145.68$ & $0.4630$ \\
  $g_4$ & $137.73$ & $1.5432$ & $s_4$ & $-137.35$ & $0.9259$ \\
\hline
   \end{tabular}
\label{tab:fredrift}
\end{table}

As we can see from Table~\ref{tab:fredrift}, the variations of $g_3$, $g_4$ and $s_2$ exceed $1.3\times10^{-7}\,2\pi\,{\rm yr}^{-1}$, which could have a great influence on the sphere of resonant action. Roughly assuming the proper frequencies of Trojans are independent of the fundamental frequency drift, we could define the boundary of the secular resonances by maximizing or minimizing $\sum_{j=2}^{8}(p_jg_j+q_js_j)$ in Eq.\,\eqref{eq:secres} with the variations listed in Table~\ref{tab:fredrift}.
Note the mean value $\bar{\nu}$ here differs a little from the frequencies in Table~\ref{tab:ffsolsys} that is calculated from a time window of 20\,Myr.
For example, the black dashed lines in the left panel of Fig.~\ref{fig:secord4} define the coverage areas of the S4\_A and G4\_F resonances in consideration of the frequency drift. Apparently, shifts of the locations of these two resonances are remarkable, especially in region ``C'', although we just give a lower limit of $\Delta\nu$. For a longer integration time, we get larger variations of the fundamental frequencies and thus the secular resonances could sweep across a broader area that is delimited by the red dashed lines. The absolute values of $g$ and $s$ in region ``C'' are much smaller than those in region ``L'' and ``R'' (cf. Fig.~\ref{fig:dynspe}), hence the locations of the secular resonances acting on the tadpole orbits are more sensitive to the frequency drift.

From Figs.~\ref{fig:secord2}--\ref{fig:secord6} we can see that there exist numerous ``C-type'' and ``G-type'' secular resonances inducing instabilities in region ``C'' and most of them are related to the fundamental frequencies in the inner Solar System. As the secular frequencies of the inner planets drift, the instability strips induced by the ``C-type'' and ``G-type'' secular resonances will sweep across the phase space in region ``C''. Finally, on timescale of 4.5\,Gyr, as long as the age of the Solar System, chaos will occupy most of the central areas, including the blue areas where the most stable Trojans reside (see the left panel in Fig.~\ref{fig:secord4} for example), and then almost no tadpole orbits will remain. In other words, there may be few primordial Earth Trojans on tadpole orbits. However, most of the horseshoe orbits with the smallest SN could survive the age of the Solar System as the ``S-type'' secular resonances excluding $s_2$ are hardly affected by the frequency drift. In view of the enormous amount of the potential extremely stable horseshoe orbits according to Fig.~\ref{fig:SNL4} (in blue), we may find abundant primordial terrestrial companions on horseshoe orbits.

\section{The Yarkovsky effect}\label{sec:yark}

\subsection{Surviving ratio}\label{subsec:sratio}

The Yarkovsky effect can produce a thermal thrust on asteroids orbiting the Sun, resulting in long term evolution of the semi-major axis for small objects with diameter $D=0.1$\,m to $\sim 40$\,km \citep{2006AREPS..34..157B}. For Earth Trojans, the Yarkovsky effect may make a great difference on their evolution due to the appropriate size and close distance to the Sun. To figure out the modification to the stability by the Yarkovsky effect, we select a sample of 986 orbits whose lifespans are longer than 1\,Gyr out from the 2870 orbits that survive in the $\sim 12$\,Myr integration to make the dynamical map (coloured points in Fig.~\ref{fig:SNL4}), and then statistically analyse their dynamical behaviour under the  Yarkovsky effect of different magnitudes that are determined by the distance to the Sun, the obliquity of the spin axis and physical properties such as the size and spin rate. As the diurnal component is much more remarkable than the seasonal component in our study, we just include the former in our computations.

According to \citet{2002Icar..157..155N}, \citet{2006AREPS..34..157B} and \citet{2013CeMDA.117...91M}, we parameterize the Earth Trojans with a bulk density of $2.5\,{\rm g/cm^3}$, a surface density of $1.5\,{\rm~g/cm^3}$, a surface conductivity of $0.001\,{\rm W/(mK)}$, an emissivity of 0.9, an albedo of 0.18 and the specific heat capacity of $680\,{\rm J/ kg/K}$.
Based on these parameters, we approximate and simplify the formula for the diurnal Yarkovsky effect developed by \citet{1999A&A...344..362V} as
\begin{equation}
  \frac{da}{dt}=3047\cdot\frac{\omega^{1/2}\cos\gamma}{R}\,,
\end{equation}
where $\omega$, $\gamma$ and $R$ represent the spin rate, obliquity of the
spin axis and radius of the asteroid respectively. The $da/dt$, $R$ and $\omega$ are in AU/Gyr, meter and $2\pi$/s respectively while $\gamma$ ranges from $0^\circ$ to $180^\circ$. For typical spin rate in the Solar System, as mentioned in \citet{2006AREPS..34..157B}, $\omega\sim{2\pi}/({5R})$, we obtain
\begin{equation}
  \label{eqn:dadtR}
  %\frac{da}{dt}=\frac{(\alpha{L})^{1/4}(KC)^{1/2}a\omega^{1/2}\cos\gamma}{3\sqrt{2}(\pi\epsilon\sigma)^{1/4}(\mu\rho)^{1/2}cR}
  \frac{da}{dt}=3415\cdot\frac{\cos\gamma}{R^{3/2}}\,.
\end{equation}

\begin{figure*}[!htbp]
	\centering
	\includegraphics[scale=0.35]{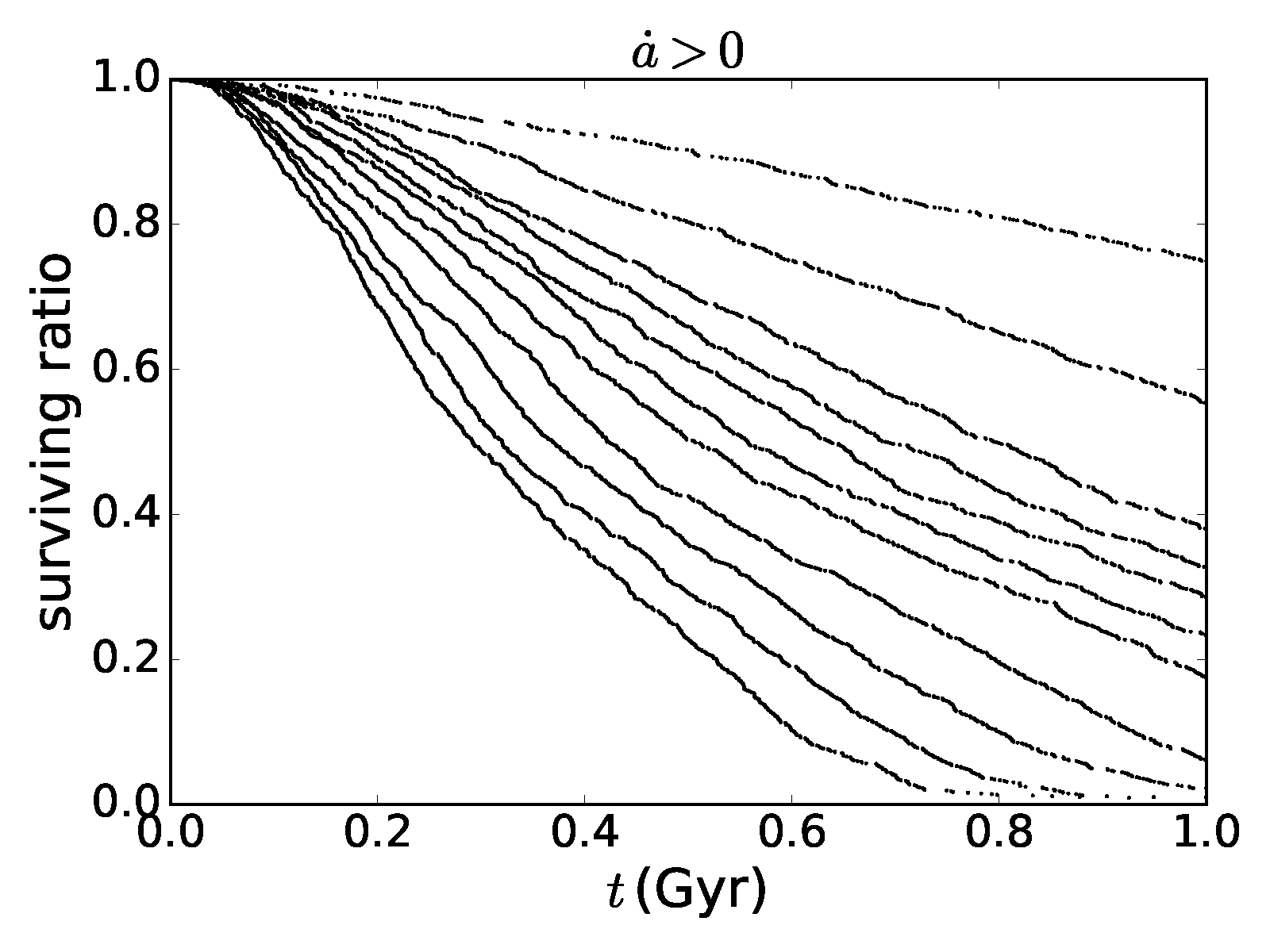}~~~~
    \includegraphics[scale=0.35]{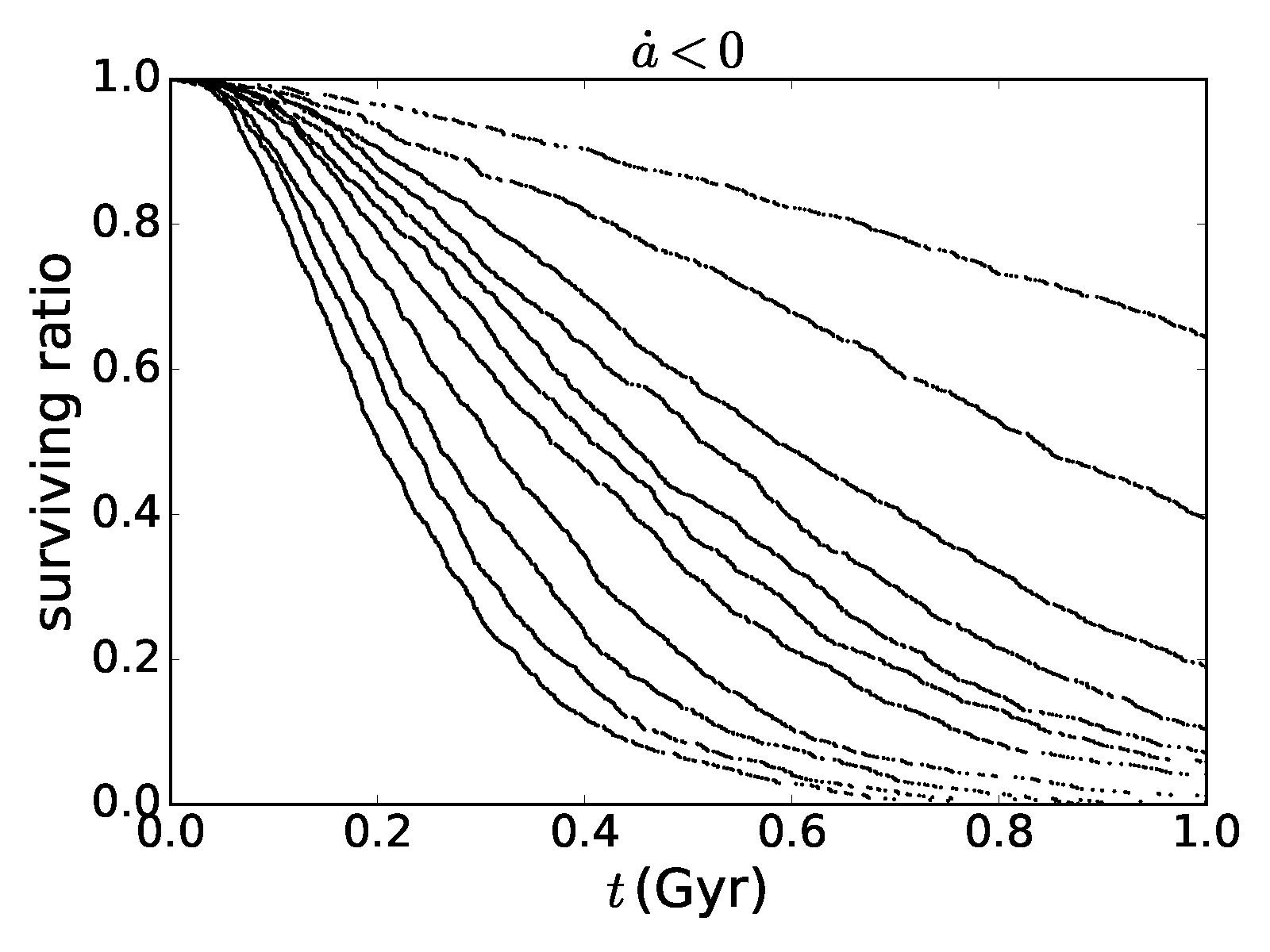}
	\includegraphics[scale=0.35]{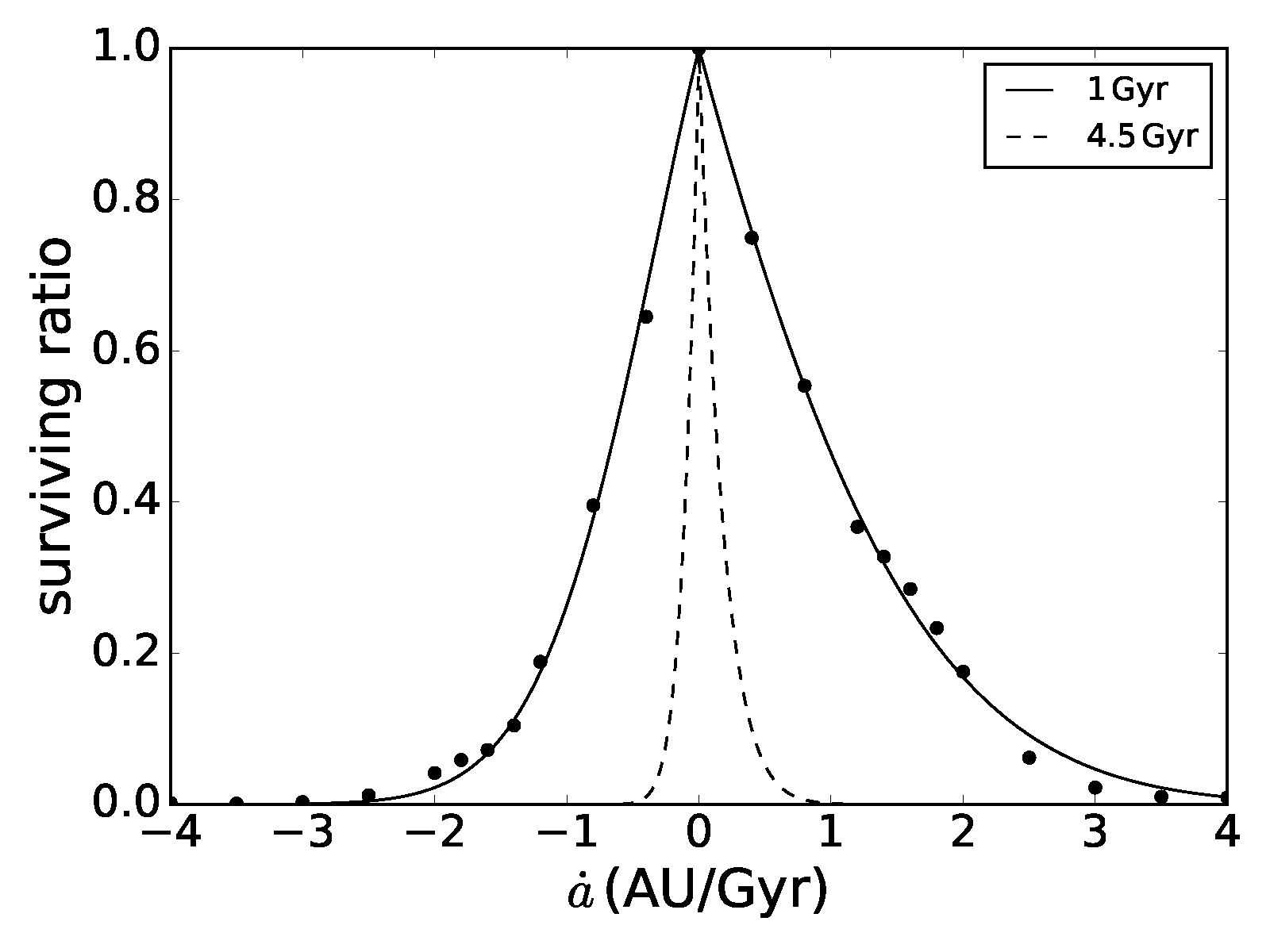}~~~~
	\includegraphics[scale=0.35]{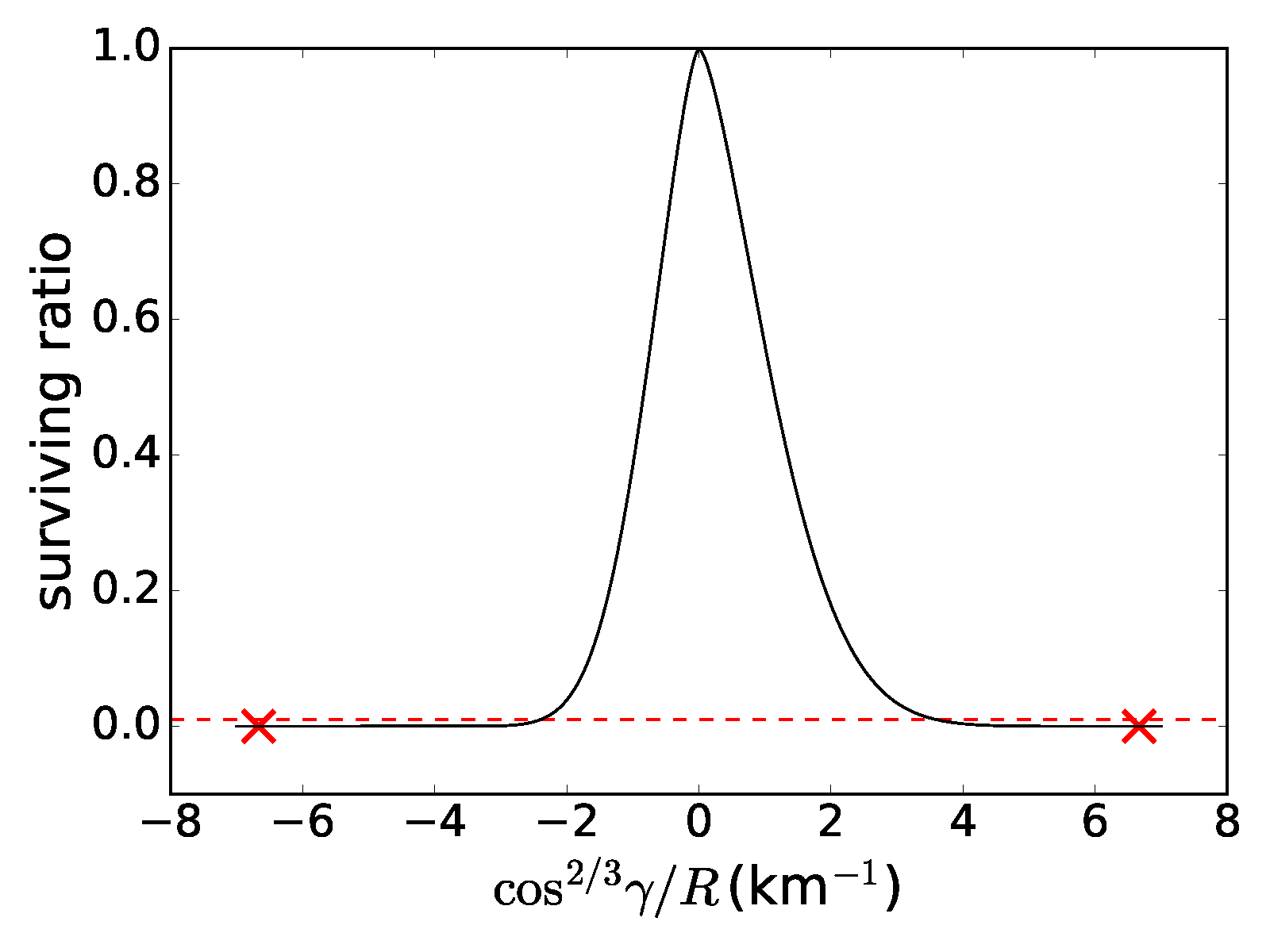}\\
	\caption{The surviving ratio against evolution time for prograde spinning Trojans ($\dot{a}>0$, upper left) and retrograde rotators ($\dot{a}<0$, upper right). In both panels, the drift rate $|\dot{a}|$ increases from 0.4 to 4\,AU/Gyr for lines from top right to bottom left. In the two lower panels, the surviving ratio at 1\,Gyr is plotted against the drift rate $\dot{a}$ (left) and $\cos^{2/3}\gamma/R$ (right). %In the lower left panel,
	The black points are numbers from the integrations and the solid curve represents the numerical fit. The dashed line stands for the surviving ratio at 4.5\,Gyr obtained by the extrapolation of Eq.\,\eqref{eqn:fitratio}. In the lower-right panel, the solid curve is transformed from the dashed line in the left panel (surviving ratio at 4.5\,Gyr) by making use of Eq.\,\eqref{eqn:dadtR}. The red crosses represent the surviving ratio for $\gamma=0^\circ,R=150$\,m and $\gamma=180^\circ,R=150$\,m. The red horizontal dashed line indicates a surviving ratio of $p=1\%$. }
	\label{fig:ratio}
\end{figure*}

As mentioned before, the diurnal Yarkovsky effect depends on a variety of physical quantities, and different combinations of these parameters may lead to the same effect, so we exert different drift rates $\dot{a}$ on Earth Trojans instead of varying physical quantities in practice to avoid the degeneracy of the specific characteristics. After some tests, we range the drift rates from $-4$ to $4$\,AU/Gyr where the negative values indicate the retrograde spin ($\cos\gamma <0$ in Eq.\,\eqref{eqn:dadtR}) and the positive ones indicate the prograde spin. It should be noted that $\dot{a}$ here is referred to the derivative of Kepler orbital element, so it can be considered as an equivalent force. We therefore exert the equivalent forces with different strengths on each of the 986 Trojan orbits, and check whether they can stay in the 1:1 MMR under the semi-major axis drift due to Yarkovsky effect. Limited by the computing resource, we integrate the system to 1\,Gyr. We count the number of survived Trojans during the integrations, compute the ratio of them to all 986 selected orbits, and present the results in Fig.~\ref{fig:ratio}.

Surely the strongest Yarkovsky effect will severely destabilize the orbits and lead to the clearance of Earth Trojans. However, the larger size and the slower rotation the Trojans are of, the weaker the Yarkovsky effect will be. The surviving ratio $p$ of affected Trojans must change with the drift rates $\dot{a}$ as well as the evolution time $t$, as the plots in  Fig.~\ref{fig:ratio} clearly show. For any given $\dot{a}$, empirically, we suppose the surviving ratio $p$ decreases with time $t$ following a rule as below
\begin{equation} \label{eqn:fitratio}
p=\exp\left[\alpha\, t^\beta\right],
\end{equation}
where $\alpha(\dot{a})$ and $\beta(\dot{a})$ are coefficients depending on $\dot{a}$. We then find that $\alpha$ and $\beta$ can be well fitted by
\begin{equation}
\begin{aligned}
  \alpha=&-\left(b_1 \dot{a}+b_2 \dot{a}^2+b_3 \dot{a}^3+b_4 \dot{a}^4\right), \\
  \beta= & c_1\exp[c_2 \dot{a}] + c_3 \exp[c_4\dot{a}],
\end{aligned}
\end{equation}
where $b_j, c_j$ $(j=1,2,3,4)$ are coefficients of the best fit. The solid curve in the lower left panel of Fig.~\ref{fig:ratio} stands for such a fitting function $p(\dot{a},t)$ for $t=1$\,Gyr, and it matches the points obtained from integrations fairly well. Encouraged by this consistence, we present also an extrapolation of Eq.\,\eqref{eqn:fitratio} to $t=4.5$\,Gyr (the dashed line), which gives an estimation of the surviving ratio of Earth Trojans at the age of the Solar System.

Combining Eqs.\,\eqref{eqn:dadtR} and \eqref{eqn:fitratio}, we get the expression of $p(\cos^{2/3}\gamma/R,t)$. Taking the size of $2010~{\rm TK_7}$ ($\sim 300$\,m in diameter) into consideration, we calculate the surviving ratio after 1\,Gyr evolution under the Yarkovsky effect and find out $p=0.034$ for the exactly retrograde rotation ($\gamma=180^\circ$) and $p=0.20$ for the exactly prograde rotation ($\gamma=0^\circ$).
At the age of 4.5\,Gyr, the surviving ratio $p$ reduces significantly. We display $p$ against $\cos^{2/3}\gamma/R$ at $t=4.5$\,Gyr in the lower right panel of Fig.~\ref{fig:ratio}. Now for Earth Trojans of 300\,m in size, we find $p=5.4\times10^{-21}$ for $\gamma=180^\circ$ and $p=2.7\times10^{-7}$ for $\gamma=0^\circ$ (red crosses in Fig.~\ref{fig:ratio}), both  practically approaching zero. In other words, the primordial Earth Trojans as large as $2010~{\rm TK_7}$ must have been driven out from the 1:1 MMR by the Yarkovsky effect unless their spin axes lie almost in the orbital plane, where the seasonal (but not diurnal) component of the Yarkovsky effect dominates.

If we take the surviving ratio of 1\% as the cut-off probability, the critical values of $\cos^{2/3}\gamma/R$ for Earth Trojans having retrograde and prograde spin after 1\,Gyr evolution are $-7.68\,{\rm km}^{-1}$ and $11.1\,{\rm km}^{-1}$, which means that the smallest radius of Earth Trojans that survive for 1\,Gyr in the 1:1 MMR is 130\,m and 90\,m, respectively for the exactly retrograde ($\gamma=180^\circ$) and prograde rotators ($\gamma=0^\circ$).
For primordial Earth Trojans, the evolution time $t=4.5$\,Gyr changes these two numbers to 400\,m and 278\,m, corresponding to absolute magnitude ${\rm H}=18.00$ and 18.79 if an albedo 0.18 is adopted. 

Of course, if other albedo is adopted, or if the spin axis of Earth Trojans is not perpendicular to the orbital plane, the critical size of Earth Trojans that survive the Yarkovsky effect could be different. Suppose three different albedos ($\alpha$), of which $\alpha=0.1$ is that of 2010 ${\rm TK_7}$ \citep{2011Natur.475..481C}, $\alpha=0.05$ for C-type asteroid and $\alpha=0.25$ for S-type asteroid, we calculate the critical absolute magnitudes of prograde and retrograde spinning Earth Trojans that can survive for 1\,Gyr and 4.5\,Gyr under the Yarkovsky effect and list them in Table~\ref{tab:hmax}.

\begin{table}[htbp]
	\caption{The upper limits of absolute magnitudes (${\rm H}$) of Earth Trojans of different albedos ($\alpha$) surviving for 1\,Gyr and 4.5\,Gyr. The subscripts `p' and 'r' indicate the prograde ($\gamma=0^\circ$) and retrograde ($\gamma=180^\circ$) spin. In the  bottom row, the range of obliquity ($\gamma$) of possible primordial Earth Trojans with absolute magnitude ${\rm H}>20.5$ is listed (see text).}
	\centering
	\begin{tabular}{c|ccc}
		\hline
		{} & $\alpha=0.05$~~ & $\alpha=0.10$~~ & $\alpha=0.25$~~ \\
		\hline
		${\rm H}(1\,{\rm Gyr})$  & $22.5_{\rm p}/21.7_{\rm r}$ & $21.8_{\rm p}/21.0_{\rm r}$ & $20.9_{\rm p}/20.1_{\rm r}$ \\
        ${\rm H}(4.5\,{\rm Gyr})$ & $20.0_{\rm p}/19.3_{\rm r}$ & $19.4_{\rm p}/18.6_{\rm r}$ & $18.4_{\rm p}/17.6_{\rm r}$ \\
		$\gamma$ & $42^\circ$--$116^\circ$ & $67^\circ$--$103^\circ$ & $76^\circ$--$98^\circ$ \\
		\hline
	\end{tabular}
	\label{tab:hmax}
\end{table}

Note that the magnitudes listed in Table~\ref{tab:hmax} are the upper limits. For example, ${\rm H}(4.5\,{\rm Gyr})=19.4_{\rm p}/18.6_{\rm r}$ for $\alpha=0.10$ means that any Earth Trojan of albedo 0.10 must be brighter than ${\rm H}=19.4/18.6$ if it is a prograde/retrograde rotator so that it is large enough in size to survive the Yarkovsky effect for the age of the Solar system. 

When the spin axis tilts away from the normal direction of the orbital plane, the Yarkovsky effect declines. Set a detection limit of Earth Trojans as ${\rm H}=20.5$ \citep{2018LPI....49.1149C}, we calculate the possible obliquity of Earth Trojan that can survive for 4.5\,Gyr and list the results in Table~\ref{tab:hmax} for reference. Again, for $\alpha=0.1$ as 2010 ${\rm TK_7}$, any primordial Earth Trojan fainter than ${\rm H}=20.5$ must have an obliquity $\gamma\in(67^\circ, 103^\circ)$. 

\subsection{Asymmetry of the stability}

As shown in Fig.~\ref{fig:ratio}, the surviving ratio displays an obvious asymmetry about $\dot{a}=0$. With the same drift rate $\dot{a}$ and at the same evolution time $t$, the orbits in prograde rotation are more likely to be stable than those in retrograde rotation. This asymmetry comes from the combined effect of the 1:1 MMR and the Yarkovsky effect.

\begin{figure*}[htbp]
	\centering
	\includegraphics[width=0.32\textwidth]{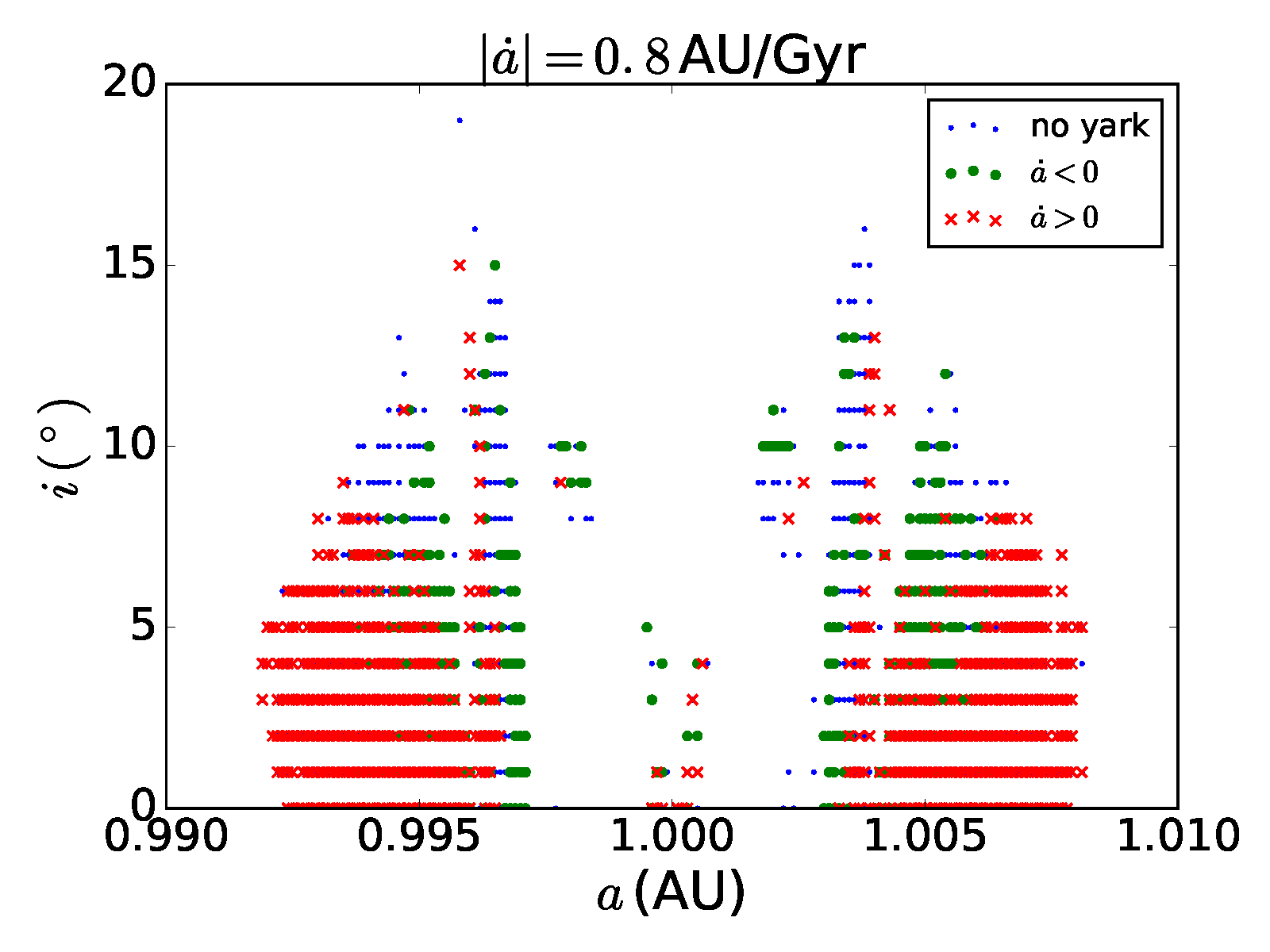}
	\includegraphics[width=0.32\textwidth]{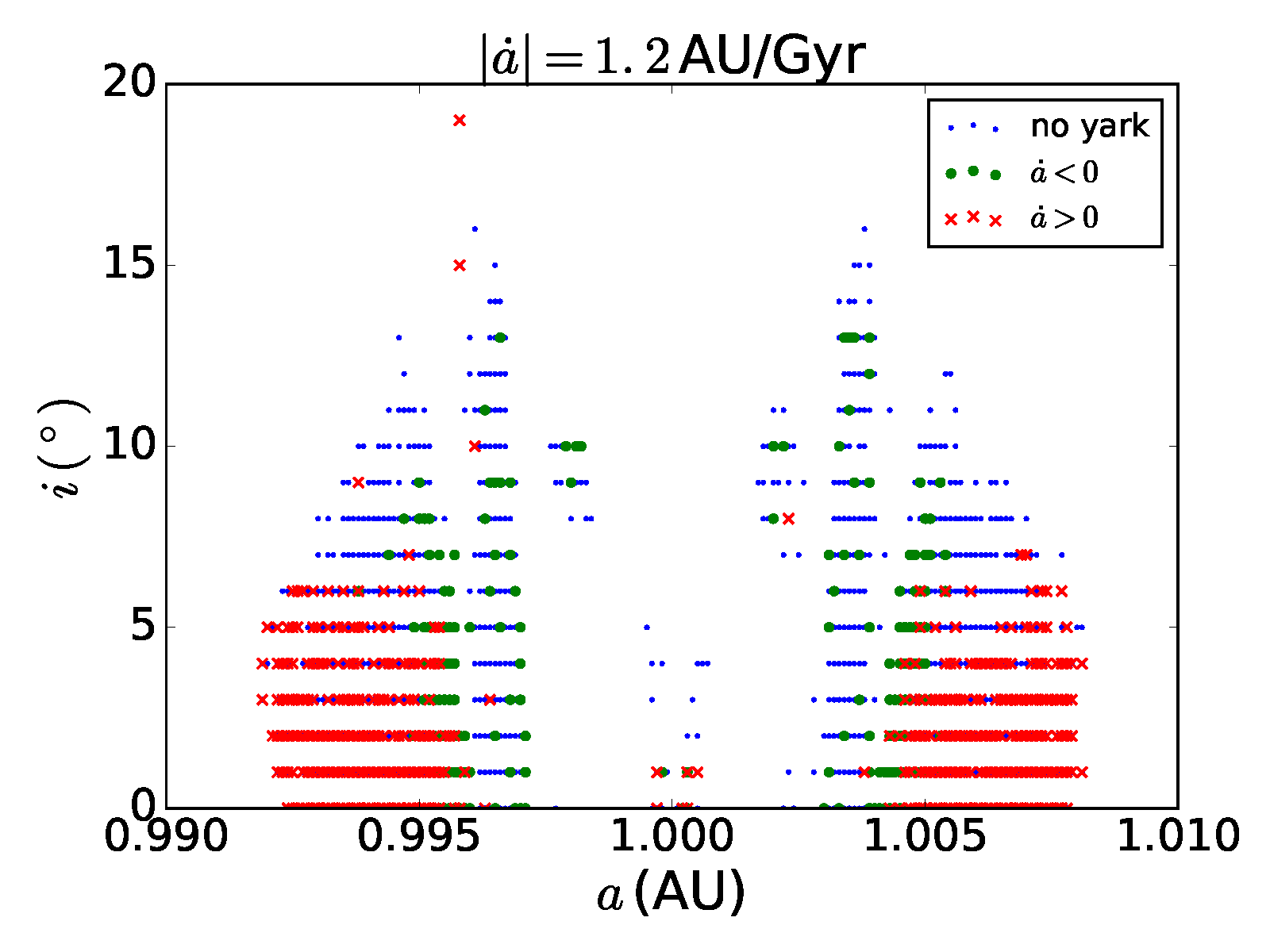}
	\includegraphics[width=0.32\textwidth]{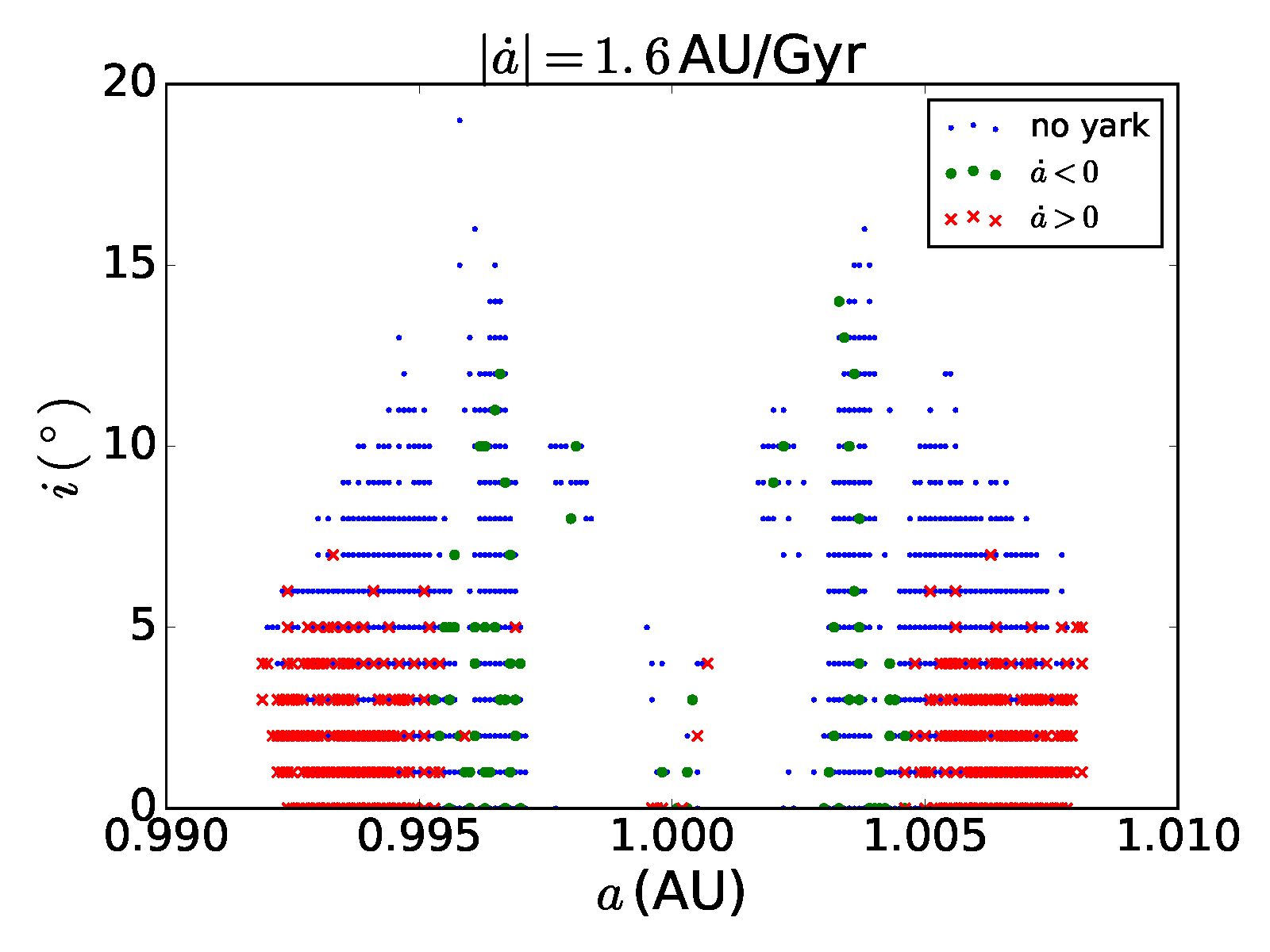}
	\caption{The distributions of initial conditions $(a_0,i_0)$ of surviving orbits after 1\,Gyr evolution. The blue points are the 986 selected orbits without the Yarkovsky effect. The green points and red crosses indicate the survivors with Yarkovsky effect in retrograde and prograde rotation, respectively. From left to right, the absolute value of semi-major axis drift rates is 0.8, 1.2 and 1.6\,AU/Gyr, respectively. Several points around $i=30^\circ$ are ignored for a better vision.}
	\label{fig:dist}
\end{figure*}

In Fig.~\ref{fig:dist}, we display on the $(a_0, i_0)$ plane the selected 986 orbits that survive the 1\,Gyr evolution. After the Yarkovsky effect of different strength is introduced, some orbits lose stability. We see clearly that the retrograde spinning survivors ($\dot{a}<0$) gather around the inner side of the stable region, separated from the prograde rotators ($\dot{a}>0$), whose rendezvous is around the outer sides. As the Yarkovsky effect gets stronger (larger drift rate $|\dot{a}|$), the number of survivors for both prograde and retrograde rotators declines, but such clustering becomes more outstanding. Actually, a longer evolution time will lead to the same appearance as the Yarkovsky force acts like a cumulative effect.

\begin{figure}[htbp]
		\centering
	\includegraphics[scale=0.21]{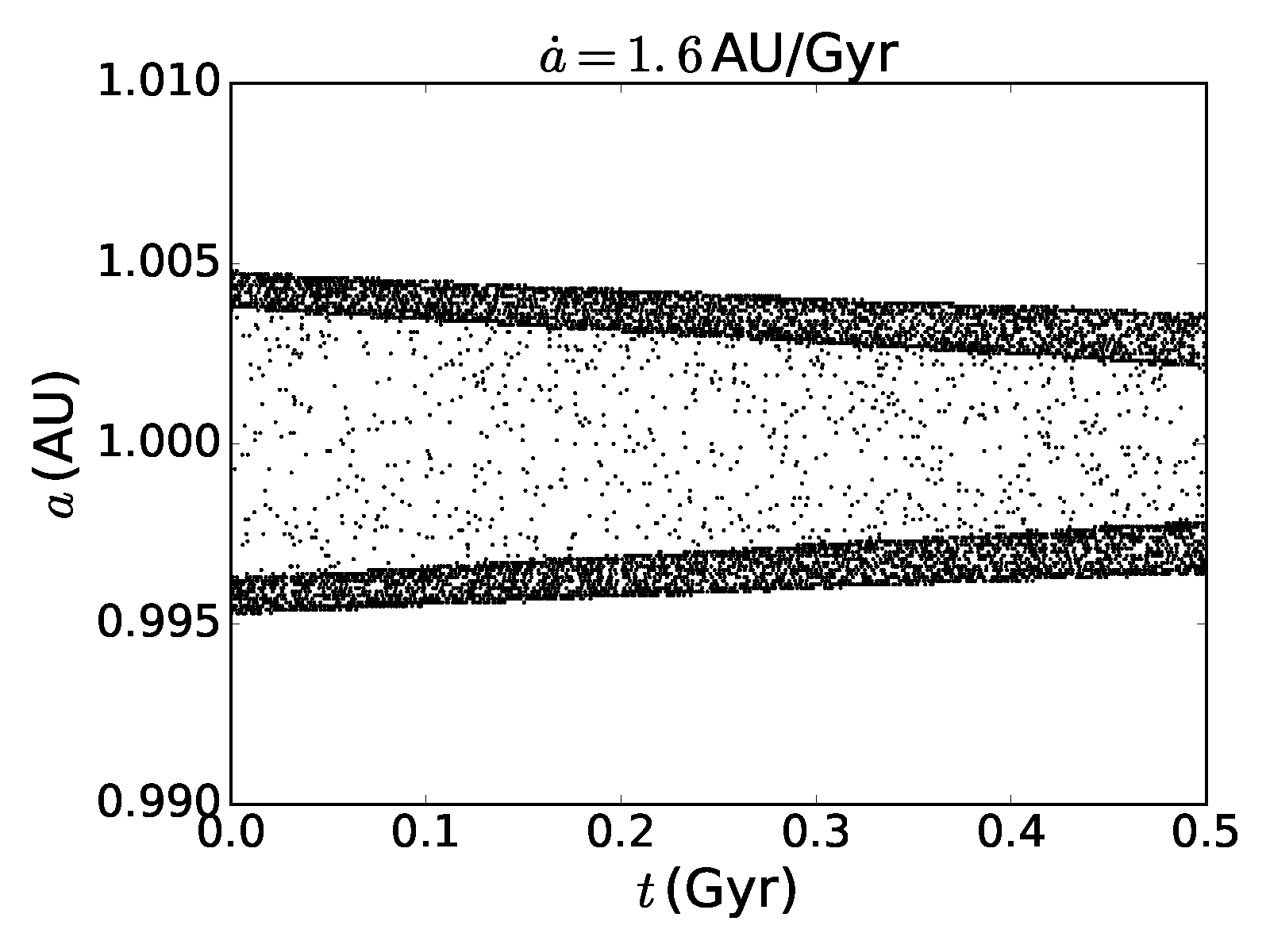}
	\includegraphics[scale=0.21]{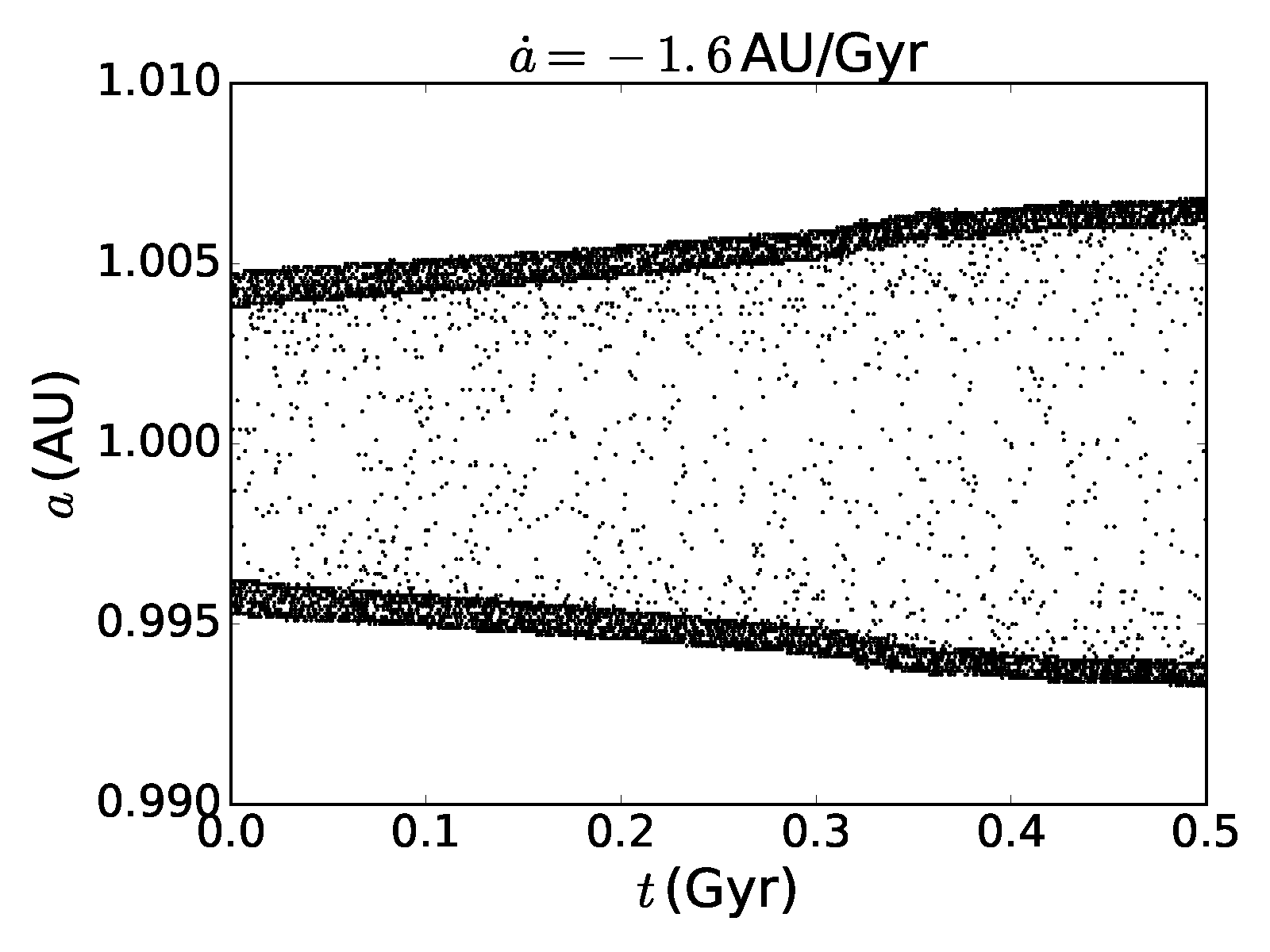}
	\caption{Time evolution of the semi-major axis for one selected orbit whose lifespan is longer than 1\,Gyr. The left panel is for a drift rate of $1.6~{\rm AU/Gyr}$ corresponding to prograde rotation while the right panel for an opposite drift due to retrograde rotation.}
	\label{fig:dayark}
\end{figure}

\citet{2017MNRAS.471..243W} proved that the libration amplitude of the semi-major axis in the 1:1 MMR undergoes remarkable variation when the Yarkovsky effect works. For prograde rotators, the amplitudes decrease with time while for retrograde rotators, the contrary is the case. As an example, we show in Fig.~\ref{fig:dayark} the variations of semi-major axes of two Earth Trojans with the same initial orbital conditions but with opposite spinning direction thus opposite Yarkovsky effect. Clearly, the Earth Trojan in prograde rotation ($\gamma=0^\circ$ thus $\dot{a}>0$) gets closer and closer to the libration centre of $a\approx 1$\,AU as time goes by. Once it crosses inward the boundary of the tadpole region, the instability caused by the secular resonances and frequency drift as described in Section \ref{sec:fma} may destabilize its orbit. Hence the closer the orbit is to the central area, the more likely it is to lose its stability. The instability in the central part of the 1:1 MMR leaves no country for Earth Trojans to hide. The same theory works for the retrograde rotators as well. The instability around the outer edge of the stability region will eliminate the orbits in expansion of the semi-major axis.

Because on the average the 986 selected orbits on the $(a_0,i_0)$ plane are further to the inner instability boundary than they are to the outer instability boundary (see Fig.~\ref{fig:dist}), \textbf{it is} easier for them to approach the outer boundary than the inner one under the semi-major axis drifts of the same rate but with opposite directions. In other words, the Earth Trojans of retrograde rotation ($\dot{a}<0$) are relatively easier to be driven out of the stable region in the 1:1 MMR. This finally makes the asymmetry of the surviving ratio in Fig.~\ref{fig:ratio}. Since the libration amplitudes of semi-major axis in the 1:1 MMR must keep increasing or decreasing for any nonzero Yarkovsky effect, the surviving ratio will tend to 0 after enough long time.%, which means there would be no Earth Trojans remained in the end.

Orbits initially outside the 1:1 MMR region may drift towards the resonance region with the help of the Yarkovsky effect. For inner orbits ($a<1$\,AU) in prograde rotation, they can approach and enter the resonance region due to a positive drift rate. At the time they cross the boundary from left side as $a$ increases, the Yarkovsky effect acts together with the resonance to reduce the libration amplitudes of the semi-major axes. Consequently, they are promoted to go through the resonance region. However, for any asteroids initially reside in the outer region ($a>1$\,AU) in retrograde rotation, at the time they cross the boundary from the right side as $a$ decreases, the Yarkovsky effect will increase the libration amplitudes of the semi-major axes, which will prevent them from entering further into the 1:1 MMR. Therefore, the Yarkovsky effect provides an effective one-way shield, preventing asteroids in the outer region ($a>1$\,AU) from entering the 1:1 MMR by virtue of the drift of the semi-major axis. This makes a bias favouring in a supplement of Earth Trojans from the inner side of the Earth orbit. In other words, if Earth Trojans are supplemented mainly by wandering objects through migration with the help of the Yarkovsky effect, most of them should have prograde rotation.

\section{Conclusion and Discussion} \label{sec:disc}
In this paper we were motivated to locate the stability regions around the Earth triangular Lagrange points $L_4$ and $L_5$. Simulations show that there are no outstanding dynamical differences between the orbits around these two points. We implemented the spectral number (SN) as the stability indicator and portrayed a much detailed dynamical map on the initial plane $(a_0,i_0)$. We found that no stable orbits appear in the areas above $i_0=37^\circ$. Two main stability regions disconnected from each other are sheltered from chaos. One of them locates at low inclinations ($i_0\la15^\circ$) while the other occupies moderate inclinations ($24^\circ\la{i_0}\la37^\circ$). The most stable orbits (${\rm SN}<100$) reside below $i_0\approx 10^\circ$ and most of them could survive the age of the Solar System. The chaos is widely distributed both inside and outside the stability regions. Furthermore, they are spreading to erode the stability regions.

The Kozai mechanism takes place in high inclination area ($i\ga 40^\circ$) for an exchange of eccentricities. The subsequent close encounters with planets scatter the Trojans away. Moreover, the overlap of the secular resonances involving $\nu_2$ and $\nu_3$, even $\nu_7$ and $\nu_8$, brings in chaos for high-inclined orbits. However, the apsidal secular resonance with Jupiter still helps hold a window around $50^\circ$ for orbits surviving for $\sim 6$\,Myr. Uranus and Neptune could make some difference to the stability in a relative short time scale.

For a further research of the resonance mechanism, we implemented a frequency analysis method to determine the locations of the secular resonances. The $\nu_3$ and $\nu_4$ secular resonances are found to be responsible for the instability gap separating the low and moderate inclination areas. They (especially $\nu_4$) have a large resonance width and could destabilize the orbits nearby by exciting their eccentricities. Most orbits undergo a small variation in inclination ($\Delta{i}\leq 10^\circ$) during their life except those surrounding the $\nu_{14}$ nodal secular resonance ($\Delta{i}\sim 20^\circ$). The transition from libration to circulation of $\Delta\Omega_3$ also varies the inclination for orbits with different $i_0$.

Higher-degree secular resonances, of which the representatives are plotted in the secular resonance maps, give rise to the fine structures in dynamical maps. The separatrices between the tadpole and horseshoe orbits at $1\pm 0.0028\,{\rm AU}$ divide the whole phase space into three regimes and the orbits around the separatrices lose their stabilities in a short time. We classify the high-degree secular resonances into three types, of which ``G-type'' only involves the apsidal precession of the Trojans while ``S-type'' only involves the nodal precession. ``C-type'' indicates the combination of the apsidal and nodal precession of the Trojans. The tadpole region in the center are mainly under the governance of the ``C-type'' and ``G-type'' secular resonances while the horseshoe regions on both sides are sculpted by the ``S-type'' resonances.

\citet{2000MNRAS.319...63T} indicated that there could be several hundred of primordial terrestrial companions provided the width of the stability zone around the Earth is $0.005\,{\rm AU}$. We find in this paper that under the influence of the frequency drift of the inner planets, the tadpole orbits can hardly survive the age of the Solar systems as the resonances they are trapped in show strong sensitivity to the precession rates of the planets. The instabilities can cover most of the tadpole cloud where the orbits with the longest lifespan reside (blue areas in the dynamical map) in a time long enough. To the contrary, most horseshoe orbits could remain almost unaffected and thus we have a chance to find the primordial terrestrial companions on horseshoe orbits.

The Yarkovsky effect modifies the stabilities of Earth Trojans by inducing the long term evolution of their semi-major axis. We obtain the expression of the surviving ratio numerically to describe how different magnitudes of the Yarkovsky effect, which depends on the obliquity of the spin axis and physical properties such as the size and spin rate, affect the stabilities of Earth Trojans. The surviving ratio displays an asymmetry in prograde and retrograde rotations. %due to the different distances to the instabilities and different variation rates of $\Delta{a}$ in the 1:1 MMR.
Driven by the Yarkovsky effect, asteroids outside the 1:1 mean motion resonance may enter the resonance region, but the Yarkovsky effect has opposite influences on the prograde and retrograde spinning asteroids. As a result, asteroids are much more likely to enter the Trojan region from the inner region ($a<1\,{\rm AU}$) and they must be in prograde rotation.

Adopting the typical values of physical and thermal parameters (bulk density, surface density, thermal conductivity, specific heat capacity, emissivity, etc) and typical spin-size relation, we find the threshold value of $\cos^{2/3}\gamma/R$ ($\gamma$ and $R$ are the obliquity and radius) for Earth Trojans to survive the Yarkovsky effect. Further taking a surviving ratio of 1\% as the critical probability, we find that an Earth Trojan surviving for 1\,Gyr must have a radius larger than 90\,m (130\,m) if it has an exactly prograde (retrograde) spin. For those surviving the Solar system age (4.5\,Gyr), the smallest radius is 278\,m for prograde spin and 400\,m for retrograde spin, respectively.

The OSIRIS-REx (Origins, Spectral Interpretation, Resource Identification, and Security-Regolith Explorer) spacecraft has been launched on 8 September 2016 to conduct an asteroid study and sample-return mission. From February 9 to February 20, 2017, the spacecraft was located near the Sun-Earth $L_4$ point to survey the sky covering $\sim 12\,~{\rm deg}^2$ for Earth Trojans. \citet{2018LPI....49.1149C} claimed that no Earth Trojans were detected from the obtained images and this sets an upper limit for the population of Earth Trojans around the $L_4$ point. Making use of the sensitivity curve, \citet{2018LPI....49.1149C} estimated that there could be no more than $73\pm 22$ Trojans around $L_4$ of absolute magnitude ${\rm H}=20.5$.

Our calculations indicate that any Earth Trojans with spin axis perpendicular to the orbital plane and albedo $0.05$ have to be brighter than ${\rm H}=20.0$ to survive the Yarkovsky effect for 4.5\,Gyr. If an albedo 0.10 as $2010~{\rm TK_7}$ is adopted, they must be brighter than ${\rm H}=19.4$. In this sense, the $2010~{\rm TK_7}$ with ${\rm H}=20.7$ cannot be a primordial Earth Trojan. Actually it is just temporarily in the Trojan orbit \citep{2011Natur.475..481C,2012A&A...541A.127D}. These estimations, together with the null detection by the OSIRIS-REx mission, make us believe that there is only a very small possibility that any primordial Earth Trojan may still exist currently. 

The Yarkovsky effect will be not strong enough to drive Earth Trojans out of the resonance within the age of the Solar system if the obliquity approaches $\gamma=90^\circ$. Our calculations show that the primordial Earth Trojans of albedo 0.10 fainter than the detection limit ${\rm H}=20.5$ must have the obliquity limited in a narrow range of $67^\circ$--$103^\circ$, which is unlikely to be common. 

Due to the uncertainties of the physical and thermal parameters of asteroids, we cannot deny absolutely the possibility of some exceptional primordial Earth Trojans existing nowadays. And of course, asteroids in the temporary Earth Trojan orbits cannot be excluded. 

\begin{acknowledgements}
We thank the anonymous referee for his comments that helped us improve the manuscript. This work has been supported by the National Natural Science Foundation of China (NSFC, Grants No.11473016, No.11473015 \& No.11333002). R.Dvorak wants to acknowledge the support from the FWF project S11607/N16.
\end{acknowledgements}

\bibliographystyle{aasjournal}
%\bibliography{mp}
%\clearpage

\end{document}